\documentclass[fleqn,usenatbib]{mnras}

\usepackage{newtxtext,newtxmath}
\usepackage[T1]{fontenc}

\DeclareRobustCommand{\VAN}[3]{#2}
\let\VANthebibliography\thebibliography
\def\thebibliography{\DeclareRobustCommand{\VAN}[3]{##3}\VANthebibliography}

\usepackage{graphicx}	% Including figure files
\usepackage{amsmath}	% Advanced maths commands

\usepackage{multirow}

\newcommand{\bham}{School of Physics and Astronomy, University of Birmingham, Birmingham B15 2TT, UK}
\newcommand{\igw}{Institute for Gravitational Wave Astronomy, University of Birmingham, Birmingham B15 2TT, UK}

\title[Bulge masses of TDE hosts]{The bulge masses of TDE host galaxies and their scaling with black hole mass}

\author[P.~Ramsden et al]{{Paige Ramsden$^{1}$}\thanks{Contact e-mail: \href{mailto:pxr754@alumni.bham.ac.uk}{pxr754@alumni.bham.ac.uk}}, Daniel Lanning$^{1}$, Matt Nicholl$^{1,2}$, Sean L.~McGee$^{1}$
\\
$^{1}$\bham\\
$^{2}$\igw
}

\usepackage{lscape}

\date{Accepted XXX. Received YYY; in original form ZZZ}

\pubyear{2021}

\begin{document}
\label{firstpage}
\pagerange{\pageref{firstpage}--\pageref{lastpage}}
\maketitle

\begin{abstract}

Tidal disruption events (TDEs) provide a means to probe the low end of the supermassive black hole (SMBH) mass distribution, as they are only observable below the Hills mass ($\lesssim 10^8 {\rm M}_{\odot}$).
Here we attempt to calibrate the scaling of SMBH mass with host galaxy bulge mass, enabling SMBH masses to be estimated for large TDE samples without the need for follow-up observations or extrapolations of relations based on high-mass samples.
We derive host galaxy masses using \textsc{prospector} fits to the UV-MIR spectral energy distributions for the hosts of 29 well-observed TDEs with BH mass estimates from \textsc{mosfit}. We then conduct detailed bulge/disk decomposition using SDSS and PanSTARRS imaging, and provide a catalog of bulge masses. We measure a positive correlation between SMBH and bulge mass for the TDE sample, with a power-law slope of 0.28 and significance $p=0.06$ (Spearmans) and $p=0.05$ (Pearsons), and an intrinsic scatter of 0.2\,dex. Applying MC resampling and bootstrapping, we find a more conservative estimate of the slope is $0.18\pm0.11$, dominated by the systematic errors from \textsc{prospector} and \textsc{mosfit}. This is shallower than the slope at high SMBH mass, which may be due to a bias in the TDE sample towards lower mass BHs that can more easily disrupt low-mass stars outside of the event horizon. When combining the TDE sample with that of the high mass regime, we find that TDEs are successful in extending the SMBH - stellar mass relationship further down the mass spectrum and provide a relationship across the full range of SMBH masses.

\end{abstract}


%%%%%%%%%%%%%%%%%%%%%%%%%%%%%%%%%%%%%%%%%%%%%%%%%%

%%%%%%%%%%%%%%%%% BODY OF PAPER %%%%%%%%%%%%%%%%%%

\section{Introduction}
\label{s:intro}

Supermassive black holes (SMBHs) exist within the nuclei of most galaxies and their properties couple with those of their hosts. The measured mass of the central SMBH correlates with the galaxy's bulge mass, luminosity and stellar velocity dispersion \citep[e.g.][]{Gebhardt2000,haring,Ferrarese2005}. These scaling relationships are thought to indicate SMBH influence on the formation and evolution of the galaxies that they lie within. If the slope can be well constrained, such relations make it possible to estimate the masses of large samples of SMBHs, enabling studies from early structure formation to the evolution of SMBHs \citep{Gultekin2009,kormendy,McConnell2013}. However, as direct SMBH measurements are most commonly available for massive active galactic nuclei (AGN) and a handful of very nearby systems, the commonly-used relationships are calibrated only by samples that are dominated by SMBHs more massive than $\gtrsim10^7$\,M$_\odot$ \citep[e.g.][]{kormendy,Bentz2013}.

However, at least one important physical phenomenon is restricted to SMBHs of lower mass: tidal disruption events (TDEs) of stars \citep{hills, rees}. If a star is unlucky enough to pass too close to a SMBH, tidal forces outweigh the binding energy of the star and so it is torn apart. This occurs at the `tidal radius',

\begin{equation}
R_T = \left(\frac{M_{\rm BH}}{M_*}\right)^{\frac{1}{3}}R_* 
\end{equation}

for a black hole of mass $M_{\rm BH}$ and a star of mass $M_*$ and radius $R_*$. If this occurs outside of the event horizon of the black hole,

\begin{equation}
R_G = \frac{2GM_{\rm BH}}{c^2},
\end{equation}

approximately half of the disrupted material falls back onto the SMBH, likely forming an accretion disk, while the other half becomes unbound \citep{rees, lacey}. Both the circularization and ultimate accretion of the bound matter can cause a luminous flare, with characteristic luminosity of $\sim 10^{44} \hbox{erg s}^{-1}$ \citep{rees,Gezari2021}. However, if the tidal radius is smaller than the gravitational radius of the SMBH, the star's orbit will intersect the event horizon before disruption, and so it will be swallowed whole, without a visible flare \citep{hills, macleod}. The requirement $R_T > R_G$ introduces a critical SMBH mass, called the Hills mass, to produce an observable a TDE. This is $\lesssim 10^8 {\rm M}_{\odot}$ for a main-sequence star of $\sim 1 {\rm M}_\odot$, or $\lesssim 10^7 {\rm M}_{\odot}$ for a more common main-sequence star of $\sim 0.1{\rm M}$, disrupted by a non-spinning SMBH. Therefore, observations of TDEs offer a new opportunity to probe lower-mass black holes, identifying those that may otherwise lie dormant and undetected, and enabling us to study their evolution at lower parts of the mass function \citep{mockler, wevers2017}. 

TDEs are now found routinely by wide-field, time-domain surveys, such as the Panoramic Survey Telescope and Rapid Response System (Pan-STARRS; \cite{kaiser}), the Zwicky Transient Facility (ZTF; \cite{bellm}), the Asteroid Terrestrial Impact Last Alert System (ATLAS; \cite{tonry}) and the All-Sky Automated Survey for Supernovae (ASASSN; \cite{kochanek}). They can be differentiated from other luminous transients by a lack of colour evolution; a long rise time and fade time relative to most supernovae (SNe); and a power-law decline from the peak, relatively consistent with the classic $t^{-5/3}$ prediction \citep{rees,Hung2018,vanvelzen2021}. Distinct spectroscopic features include a colour temperature of $10,000-50,000$\,K and broad emission lines of H\,I, He\,II and N\,III \citep{gezari,arcavi,Blagorodnova2019,vanvelzen2021}. Most recently, 17 new TDEs were discovered by ZTF and studied by \cite{vanvelzen2021}. With the upcoming Vera Rubin Observatory and its Legacy Survey of Space and Time (LSST; \citealt{lsst}), and the Extended Roentgen Survey with an Imaging Telescope Array (eROSITA; \citealt{khabibullin, sazonov}), it is predicted that this field will be further revolutionised by the detection of thousands of TDEs per year \citep{stone, bricman, khabibullin, sazonov}.

The most precise BH mass estimates require the orbital velocity distribution of stars and/or gas close to the galactic nucleus \citep{kormendy}, which can only be measured spectroscopically. It will not be possible to obtain spectroscopic BH mass measurements for such a large sample of TDEs, yet such masses would be extremely valuable in understanding TDE properties, and in building a picture of SMBHs at the low end of their mass function.  Assuming the TDE luminosity is `prompt', i.e.~its evolution follows the fallback rate of material \citep{guillochon}, the TDE light curve can be modelled directly to derive dynamical quantities of the encounter, including SMBH mass \citep{mockler, stone}. Yet even this method requires high-cadence, multiwavelength follow-up of each TDE, due to the limited cadence of LSST and the fact that TDEs emit most of their energy in the UV. This method too becomes impractical for studying $\sim1000$ TDEs.

However, the majority of future TDEs will have excellent host galaxy imaging from their discovering surveys: LSST will achieve final depths of $\sim26-27$\,mag in stacked images. If the BH masses in TDE host galaxies correlate with properties that can be measured photometrically, constructing the TDE BH mass function can be done essentially "for free". It is known that SMBH masses correlate with the luminosity, or equivalently the mass, of their host galaxies, and in particular correlate with the mass of the inner bulge at a significance comparable to their correlation with velocity dispersion \citep{Gultekin2009,kormendy,McConnell2013}. However, this relation has been calibrated mainly at BH masses too large to produce a TDE \citep[e.g.][]{wevers19}, so it has been unclear if it applies, or differs, in the TDE regime $M_{\rm BH}\sim 10^6\,M_\odot$. 

Moreoever, TDEs often occur in unusual galaxies, where the slope of a BH--galaxy correlation could differ. They are over-represented in quiescent Balmer-strong galaxies, otherwise called post-starburst or E+A galaxies. These galaxies are defined by distinct Balmer line absorption and weak, or no, emission lines \citep{french, arcavi, graur, law-smith}. This is suggestive of low levels of current star-formation (the lack of emission lines), but with substantial activity in the last $\sim$Gyr (Balmer absorption from A-type stars) \citep{french, zabludoff}. \citet{french} found that quiescent Balmer-strong galaxies host more than one third of the observed sample of TDEs, yet they make up only 0.2$\%$ of local galaxies. It is possible that these are post-merger galaxies \citep{zabludoff}, resulting in an enhanced TDE rate caused by the interaction of two SMBHs \citep{chen}. Enhancement of the TDE rate could also be caused by the formation of quiescent Balmer-strong galaxies via galaxy interaction, unique stellar populations existing in the cores of the galaxies \citep{mockler_2022,arcavi}, or by high central stellar densities \citep{french_2020,graur}.

Armed with the growing sample of known TDEs and an ever increasing discovery rate, now is the time to confirm whether TDEs can pick out low mass SMBHs, identify whether BH masses obtained from light curve fits are reliable, and extend the calibration of SMBH-host relations to the TDE regime.
In this study, we derive host galaxy bulge masses for a sample of 29 TDEs using pre-flare photometry from optical, UV and IR sky surveys, in combination with the stellar population synthesis code \textsc{Prospector} \citep{leja}, and bulge-to-total light (B/T) ratios measured by decomposing SDSS and PanSTARRS host galaxy images. 

SMBH masses for these events have been estimated using \textsc{mosfit} (companion paper by \citealt{nicholl2022}), and we use these to derive the SMBH--bulge mass relation for the TDE sample. Combining these data with the well-studied high-mass end of the SMBH mass function, we attempt to measure the slope of the SMBH--stellar mass relation from $\lesssim10^6$\,M$_\odot$ to $\gtrsim10^{10}$\,M$_\odot$. We find a moderately significant correlation within the TDE host population, but compared to the high-mass sample it is weaker, with larger intrinsic scatter and a flatter slope. This may be the result of systematic difficulties in measuring their SMBH masses, but could also reflect the bias towards low-mass SMBHs due to event horizon suppression. 

In Section \ref{s:spectral}, we present the selection of TDE candidates and their host galaxies, alongside a description of the \textsc{Prospector} model used and choice of priors. Results for galaxy mass are displayed and compared with previous literature, and B/T ratios are used to find the galaxy bulge masses. In Section \ref{s:bh-bulge} the BH--bulge mass for the TDE sample is analysed and results are reported with their significance. Finally, the TDE sample is combined with that of the high mass regime \citep{kormendy} and the SMBH--bulge mass relationship is derived for the whole range of SMBH masses. The paper ends with discussion and conclusions in Section \ref{s:discussion}.

\section{Spectral energy distribution fits and bulge masses}
\label{s:spectral}
% Input data sample

\hypersetup{citecolor=black}
\begin{table*}
\small
\caption{\small Our sample of 29 TDEs with their locations, redshifts and the derived host galaxy bulge masses.}
\label{17_sample}
        \begin{center}
\begin{tabular}{l l l l l l l} 
 \hline\hline
 IAU Name/Discovery Name & RA & Decl. & Redshift (z)  & Classification reference & $\log(M_{\rm bulge} / M_{\odot})$ & $\log(M_{\rm BH} / M_{\odot})$ $^a$ \\ 
 \hline
 ASASSN-14ae & 11:08:40.12 & +34:05:52.23 & 0.043 & \cite{holoien} & 9.73$^{+0.15}_{-0.17}$ & 6.13$^{+0.05}_{-0.04}$ \\
 ASASSN-14li & 12:48:15.23 & +17:46:26.44 & 0.021 & \cite{holoien_class2} & 9.78$^{+0.07}_{-0.07}$ & 7.00$^{+0.08}_{-0.11}$ \\
 ASASSN-15oi & 20:39:09.18 & - 30:45:20.10 & 0.020 & \cite{holoien_class3} & 8.91$^{+0.10}_{-0.08}$ & 6.73$^{+0.02}_{-0.02}$\\
 PS17dhz/AT2017eqx & 22:26:48.30 & +17:08:52.40 & 0.109 & \cite{nicholl_class2} & 9.51$^{+0.11}_{-0.14}$ & 6.56$^{+0.08}_{-0.09}$ \\
 AT2018hco & 01:07:33.635 & +23:28:34.28 & 0.090 & \cite{van_class1} & 9.99$^{+0.09}_{-0.10}$ & 6.64$^{+0.14}_{-0.15}$ \\
 AT2018hyz & 10:06:50.871 & +01:41:34.08 & 0.046 &  \cite{dong_class} & 9.15$^{+0.09}_{-0.11}$ & 6.57$^{+0.04}_{-0.04}$ \\
 AT2018iih & 17:28:03.930 & +30:41:31.42 & 0.212 &  \cite{vanvelzen2021} & 10.54$^{+0.10}_{-0.15}$ & 6.92$^{+0.02}_{-0.02}$\\
 AT2018lna & 07:03:18.649 & +23:01:44.70 & 0.091 &  \cite{van_class2} & 9.36$^{+0.13}_{-0.14}$ & 6.67$^{+0.13}_{-0.12}$ \\
 AT2018zr & 07:56:54.530 & +34:15:43.61 & 0.071 & \cite{tucker_class} & 9.95$^{+0.10}_{-0.10}$ & 6.79$^{+0.04}_{-0.04}$ \\
 AT2019azh & 08:13:16.945 & +22:38:54.03 & 0.022 &  \cite{van_class3} & 10.11$^{+0.08}_{-0.07}$ & 6.70$^{+0.06}_{-0.07}$ \\
 AT2019bhf & 15:09:15.975 & +16:14:22.52 & 0.121 &  \cite{vanvelzen2021} & 10.28$^{+0.12}_{-0.15}$ & 6.57$^{+0.13}_{-0.12}$ \\
 AT2019cho & 12:55:09.210 & +49:31:09.93 & 0.193 &  \cite{vanvelzen2021} & 9.87$^{+0.13}_{-0.20}$ & 6.71$^{+0.09}_{-0.08}$ \\
 AT2019dsg & 20:57:02.974 & +14:12:15.86 & 0.051 &  \cite{nicholl_class} & 10.07$^{+0.10}_{-0.11}$ & 6.57$^{+0.21}_{-0.16}$ \\
 AT2019ehz & 14:09:41.880 & +55:29:28.10 & 0.074 &  \cite{gezari_class2} & 9.73$^{+0.12}_{-0.17}$ & 6.34$^{+0.04}_{-0.04}$\\
 AT2019eve & 11:28:49.650 & +15:40:22.30 & 0.064 &  \cite{vanvelzen2021} & 9.25$^{+0.09}_{-0.10}$ & 5.79$^{+0.08}_{-0.07}$ \\
 AT2019lwu & 23:11:12.305 & - 01:00:10.71 & 0.117 & \cite{vanvelzen2021} & 9.58$^{+0.09}_{-0.09}$ & 6.31$^{+0.13}_{-0.11}$ \\
 AT2019meg & 18:45:16.180 & +44:26:19.21 & 0.152 & \cite{van_class4} & 10.14$^{+0.12}_{-0.27}$ & 6.52$^{+0.06}_{-0.06}$ \\
 AT2019mha & 16:16:27.799 & +56:25:56.29 & 0.148 & \cite{vanvelzen2021} & 9.75$^{+0.08}_{-0.08}$ & 6.25$^{+0.12}_{-0.15}$\\
 AT2019qiz & 04:46:37.880 & - 10:13:34.90 & 0.015 & \cite{siebert_class} & 9.47$^{+0.10}_{-0.12}$ & 6.22$^{+0.04}_{-0.04}$ \\
 GALEX-D1-9 & 02:25:17.00 & - 04:32:59.00 & 0.326 & \cite{gezari_class3} & 10.40$^{+0.11}_{-0.11}$ & 6.79$^{+0.25}_{-0.33}$ \\
 GALEX-D3-13 & 14:19:29.81 & +52:52:06.37 & 0.370 & \cite{gezari_class4} & 10.99$^{+0.17}_{-0.16}$ & 7.00$^{+0.21}_{-0.22}$ \\
 PS1-10jh & 16:09:28.28 & +53:40:23.99 & 0.170 & \cite{gezari} & 9.64$^{+0.12}_{-0.18}$ & 7.00$^{+0.04}_{-0.06}$ \\
 PS1-11af & 09:57:26.82 & +03:14:00.94 & 0.405 & \cite{charnock_class} & 10.07$^{+0.16}_{-0.13}$ & 6.45$^{+0.05}_{-0.04}$ \\
 PTF-09djl & 16:33:55.97 & +30:14:16.65 & 0.184 & \cite{arcavi} & 10.31$^{+0.13}_{-0.15}$ & 6.42$^{+0.11}_{-0.11}$ \\
 PTF-09ge & 14:57:03.18 & +49:36:40.97 & 0.064 & \cite{arcavi} & 10.08$^{+0.10}_{-0.10}$ & 6.47$^{+0.04}_{-0.04}$ \\
 SDSS-TDE1 & 23:42:01.41 & +01:06:29.30 & 0.136 & \cite{van_class5} & 10.14$^{+0.12}_{-0.19}$ & 6.84$^{+0.31}_{-0.34}$ \\
 SDSS-TDE2 & 23:23:48.62 & - 01:08:10.34 & 0.252 & \cite{van_class5} & 10.40$^{+0.12}_{-0.16}$ & 6.66$^{+0.34}_{-0.47}$ \\
 iPTF-16axa & 17:03:34.34 & +30:35:36.60 & 0.108 & \cite{hung} & 10.07$^{+0.11}_{-0.11}$ & 7.29$^{+0.11}_{-0.13}$ \\
 iPTF-16fnl & 00:29:57.01 & +32:53:37.24 & 0.016 & \cite{blagorodnova} & 9.81$^{+0.05}_{-0.07}$ & 5.90$^{+0.15}_{-0.06}$ \\
 \hline\hline
\end{tabular}\\
$^a$ From \citet{nicholl2022}
\label{t:sample}
\end{center}
\end{table*}

\hypersetup{citecolor=blue}

\subsection{Data Sample}
\label{s:data}

The sample of TDEs were selected based on the work of \cite{vanvelzen2021}, resulting in an up to date (2020) initial sample of 39 well observed events from recent ZTF data and previous literature. Of these, 32 have BH mass measurements from \citet{nicholl2022} using the \textsc{Mosfit} TDE model\footnote{BH mass estimates for many of these events are available elsewhere in the literature \citep[e.g.][]{wevers19}, however using any galaxy-based estimates would make our analysis circular.} \citep{mockler}. To derive host galaxy masses, we require (at minimum) multicolour optical imaging of the host, either from a previous study of the TDE in question or from a wide-field survey like PanSTARRS or SDSS. Three southern hemisphere TDEs (OGLE16aaa, ASASSN-18ul/AT2018fyk, ASASSN-18pg/AT2018dyb) did not pass this cut. The final sample utilised 15 (of 17) ZTF TDEs and 14 (of 22) literature TDEs. These are listed in Table \ref{t:sample}. 

Broadband photometry of TDE host galaxies, prior to the point of disruption, was obtained through survey catalogue searches and collated to construct spectral energy distributions (SEDs). Optical data in the $u$, $g$, $r$, $i$, $z$, $y$\footnote{$u$ bands from SDSS, $y$ bands from PanSTARRS} bands were collected from SDSS and PanSTARRS. From SDSS, the default `model Mags' were used, while aperture magnitudes were retrieved from PanSTARRS. Near-infrared (NIR) and mid-infrared (MIR) data were collected from the 2MASS and WISE catalogues, via the NASA/IPAC Infrared Science Archive, whilst UV data was collected from the GALEX catalogue via VizieR. From 2MASS, the extended profile-fit magnitudes ($m_{ext}$) were retrieved for $J$, $H$ and $K_s$ bands. NUV and FUV bands were obtained from GALEX where possible.  The default WISE photometry (instrumental standard aperture magnitude) was used for all hosts, except those that are larger in projection than the default 8.25 arcsecond aperture. This occurred most commonly for TDEs at redshifts of $z \sim 0.02$. In this case we used WISE galaxy-fitting photometry if available, otherwise larger apertures were queried. All photometric data have been homogenised within the AB magnitude system.

Due to the lower resolution of WISE, issues with blending were identified. Any data that appeared to be anomalously bright in the WISE bands was investigated using higher resolution optical images from PanSTARRS or SDSS. If two or more sources were identified within the image it was assumed that blending had occured and the WISE data was removed. Blending was not considered an issue for optical/NIR bands due to much smaller point spread functions (PSFs), and was not an issue for GALEX due to a reduced number of UV-bright sources compared to IR.

\begin{table*}
\small
\begin{centering}
\caption{\small Statistical values obtained via data analysis for both the isolated TDE sample and combined sample. Values resulting from systematic errors added in quadrature are also shown. Resampling refers to the bootstrapping + perturbation method detailed in section \ref{s:analysis} \textbf{Note:} any zero values shown represent p-value (and errors) that tend to zero. These values are $\leq 0.05$.}
\label{t:stats}
\begin{tabular}{p{2cm} | p{3cm} | p{1.3cm}  | p{1.3cm} | p{1.3cm} | p{1.3cm} | p{1.3cm} | p{1.3cm} |}
\hline\hline
\multicolumn{2}{|c|}{} & Pearson Coeff ($r_p$) & Pearson\newline P-val ($P_p$) & Spearman Coeff ($r_s$) & Spearman P-val ($P_s$) & Gradient & y-intercept \\
\hline
\multirow{3}{=}{TDE sample} & {Simple fit} &   0.38 &
 0.05 &
 0.35 &
 0.06 &
 0.28 &  3.82 \\ 
%\cline{2-9}
& \vfill {Resampling} &  \vfill $0.34^{+0.08}_{-0.00}$ & \vfill $0.08^{+0.11}_{-0.05}$ & \vfill $0.33^{+0.09}_{-0.09}$ & \vfill $0.13^{+0.10}_{-0.10}$ & \vfill $0.26^{+0.07}_{-0.07}$ & \vfill $4.00^{+0.66}_{-0.67}$ \\ 
& \vfill {Resampling+systematic} &  \vfill $0.24^{+0.14}_{-0.00}$ & \vfill $0.20^{+0.38}_{-0.16}$ & \vfill $0.23^{+0.15}_{-0.15}$ & \vfill $0.31^{+0.33}_{-0.27}$ & \vfill $0.18^{+0.11}_{-0.11}$ & \vfill $4.83^{+1.04}_{-1.07}$ \\ 
% \cline{3-9} 

\hline\hline
\multirow{3}{=}{Combined TDE + high mass sample} & {Simple fit} &   0.86 &
 $<0.001$ &
 0.82 &
 $<0.001$ &
 1.34 &  $-$6.33 \\ 
%\cline{2-9}
& \vfill {Resampling} &   \vfill $0.84^{+0.01}_{-0.00}$ & \vfill
$<0.001$ & \vfill
$0.81^{+0.01}_{-0.01}$ & \vfill
$<0.001$ & \vfill
$1.31^{+0.03}_{-0.03}$ & \vfill $-6.06^{+0.31}_{-0.31}$ \\ 
& \vfill {Resampling+systematic} &  \vfill $0.82^{+0.02}_{-0.00}$ & \vfill
$<0.001$ & \vfill
$0.79^{+0.02}_{-0.02}$ & \vfill
$<0.001$ & \vfill
$1.24^{+0.05}_{-0.05}$ & \vfill $-5.33^{+0.53}_{-0.52}$ \\ 
% \cline{3-9} 

\hline\hline
\end{tabular}
\end{centering}
\end{table*}

\subsection{SED Fits}
\label{s:sed}
% 
% Description of prospector model and choice of priors - non-parametric SFH

Modelling of the SED of each galaxy was conducted using \textsc{Prospector} \citep{leja}, with all fits run on the University of Birmingham's \textsc{bluebear} high performance computing cluster. \textsc{Prospector} uses the Flexible Stellar Population Synthesis (\textsc{FSPS}) package \citep{conroy} and creates synthetic spectra for stellar populations. The modelled SED is the total of a galaxy's content: stars and gas. Within this work, we adopted the same model used by  \cite{nicholl} to analyse the host of the TDE AT2019qiz, which is an implementation of Prospector-$\alpha$ \citep{leja}. The model free parameters used were stellar mass, defined as the mass of existing stars and stellar remnants \citep{leja}; stellar metallicity; a six component star-formation history (SFH); and three parameters which control the dust fraction and reprocessing. 

\cite{vanvelzen2021} previously modelled the SEDs of the same sample of galaxies, also using \textsc{Prospector}, but included only optical and UV data and assumed a parametric SFH. It is important to note that, in this paper, a non-parametric SFH is used, which enables a better understanding of the age of the system \citep{leja}. Furthermore, the stellar population age is degenerate with mass and metallicity, hence a more accurate constraint on this will allow for a more reliable extraction of host galaxy mass \citep{leja}. The number of SFH bins, as well as size and location, has to be balanced between minimising computational resources and maximising amount of SFH information obtained. As described in \cite{leja}, we used a six component SFH which includes the current star-formation rate (sSFR) and five equal-mass bins for SFH. 

Use of the stellar metallicity free parameter is important as it plays a key role in derivation of ages, dust attenuation and masses, due to its influence on the optical-to-NIR ratio \citep{bell, mitchell}. The final three dust parameters relate to the optical depth from general interstellar dust within the galaxy and how this depth varies with wavelength, plus the optical depth of any additional dust in star-forming regions as a fraction of the average. Further discussion of these parameters can be found in \cite{leja}. In this paper, we focus on the results for stellar mass and marginalise over all other parameters; analysis of other derived properties, in particular the SFHs, will be the subject of a follow-up study.

\textsc{Prospector} produces well-sampled posterior probability distributions for each of the detailed model parameters using Markov chain Monte Carlo (MCMC) analysis, via the \textsc{emcee} package \citep{foreman}. This also allows us to marginalise over the various model parameters in order to obtain reliable mass measurements with realistic error bars that account for degeneracies between parameters. All 29 SEDs are presented in Appendix \ref{a:A}. All derived free parameters, including the galaxy mass (and bulge ratio described in section \ref{s:bulge}), are given in Appendix \ref{a:B}. The final galaxy bulge masses are presented in Table \ref{t:sample}. For the TDE sample, we find  bulge masses in the range of $10^{8.91} - 10^{10.99} M_{\odot}$, with a median of $10^{9.95} M_{\odot}$. 

The galaxy masses measured are comparable with previous literature. We find that our well-sampled total stellar mass values are systematically, slightly larger than previous measurements. In \cite{graur}, galaxy stellar masses were computed with the \textsc{LePhare} code \citep{ilbert} and SDSS ugriz cmodel magnitudes \citep{stoughton}. We found that, when fitting SEDs containing only archival photometry from SDSS our data was in complete agreement with \cite{graur}, whilst when fitting SEDs with the full wavelength range from UV to NIR/MIR our results were typically larger. Similarly, \cite{vanvelzen2018} used photometric data from SDSS and 2MASS to estimate stellar mass via \textsc{kcorrect} \citep{blanton}. Again we find that our data results in larger total stellar masses. Even more relevant, a comparison to \cite{vanvelzen2021} indicates the same result. As previously mentioned, \cite{vanvelzen2021} used only optical and UV data, primarily from SDSS, and assumed a parametric SFH, but they also used the \textsc{Prospector} code. When comparing results on host galaxy total stellar mass, we find that while values generally agree within 1-$\sigma$, on average our results are systematically larger, typically by $\sim0.3$ dex. In combination, these comparisons suggest that previous measurements may have been underestimates due to a reduced amount of archival photometry used.

\subsection{Bulge decomposition}
\label{s:bulge}

The decomposition of TDE host galaxy light into both a bulge and disk component was done using \textsc{profit} \citep{Robotham2017}. This $R$ package models the two-dimensional photometric profile of a galaxy within a Bayesian framework and therefore produces well-justified model uncertainties. The galaxy profiles were determined using $g$-band images from both the SDSS and PanSTARRS surveys.  For each image, a Moffat function \citep{Moffat1969} was fit to the nearest 10 unsaturated bright stars, the result of which was used as the point spread function for the fitted galaxy light profile. The Moffat function is commonly used as an analytic approximation to the point spread function, and contains the Gaussian function as a limiting case, but allows for the possibility of broader wings in the distribution. Nonetheless, our results are not sensitive to the choice of the functional form of the PSF. 

For each image, a segmentation map was made using \textsc{sextractor} \citep{Bertin1996}. These segmentation maps are used by \textsc{profit} during the modelling process to avoid contributions of light profiles from nearby sources and to properly compute the noise level in the image. With \textsc{profit}, we modelled the galaxy light as the combination of a de Vaucouleur bulge component (Sersic index n=4) and an exponential disk component (Sersic index n=1). For each of the bulge and the disk components, there were 6 fitted parameters: x and y central coordinate, magnitude, effective radius, orientation of the major axis, and axial ratio (minor-axis/major-axis). These components were fitted in log space with uniform priors using the Bayesian optimisation scheme from \textsc{LaplacesDemon} with the maximum likelihood galaxy model solution from the BFGS algorithm \citep{Fletcher1970} as an initial starting point. Ultimately, the bulge-to-total ratio is the ratio of the computed bulge magnitude to the total (bulge+disk) magnitude and the results are listed in Table \ref{t:big}, Appendix \ref{a:B}. In 24 cases, the B/T was computed from the SDSS $g$ image, and for the remaining 5 cases from the PanSTARRS $g$ image. When both images were available, the computed B/T agreed very well (B/T$_{\rm{SDSS}}$ - B/T$_{\rm{PanSTARRS}}$ = 0.034 $\pm$ 0.072). 

We note that \cite{french_2020} modelled four TDE host galaxies (ASASSN14li, ASASSN14ae, iPTF15af, PTF09ge) with various surface brightness profiles, and found that the disk + de Vaucouleurs profile may not be the best model in two out of their four fits. In order to test the robustness of our disk + de Vaucouleurs fits for the galaxy and bulge, we refit these four events while allowing the Sersic index of the bulge component to vary within the wide range of $n = 2$ to $n = 20$. No significant difference was seen in the B/T ratio compared with our results with the disk + de Vaucouleurs profile, with any variation well within the $1\sigma$ uncertainties.

\section{The BH - bulge mass relation}
\label{s:bh-bulge}

\subsection{Data Analysis Methods}
\label{s:analysis}

In order to identify correlations between our data, we utilised both the Pearson Correlation Coefficient $r_p$ and the Spearman Rank Correlation Coefficient $r_s$, testing the strength of linear and monotonic relationships respectively. The slope of an assumed power-law relation \citep[e.g.][]{kormendy} was determined through the best-fit straight line in $\log M_{\rm BH}-\log M_{\rm bulge}$, and was calculated using standard $\chi^2$ minimisation.

Monte Carlo (MC) sampling techniques were used to account for limitations within our data sets. To ensure any correlation was not driven by a few extreme values, we first applied a bootstrapping method. Within each iteration, we randomly draw $N$ = 29 data points with replacement, (ie, $N$ = the number of data points within the sample). These were then resampled within their errors by approximating each drawn point as a 2D Gaussian distribution using the $1\sigma$ widths of the \textsc{Prospector} posteriors and \textsc{Mosfit} posteriors. When including systematic errors in the analysis, these widths included a systematic error of 0.32 dex in galaxy mass, taken from the average difference between our masses and those for the same objects in \citet{vanvelzen2021} and \citet{graur}, and a systematic error of 0.20 in the SMBH mass \citep[from][]{mockler}, respectively. This approach, termed `bootstrapping + perturbation' by \citet{curran}, captures both the size of the errors and the effects of a finite sample size. 

The best fit gradients and intercepts, the statistical coefficients $r_p$ and $r_s$ and their p-values were calculated at each step. Each posterior was explored in full over 50,000 iterations, and probability distributions for each of the 6 statistical values were produced. Table \ref{t:stats} presents all values obtained via statistical analysis methods described, with and without including the estimated systematic errors.

\subsection{TDE Sample}
\label{s:tdesamp}

\begin{figure}
\centering
\includegraphics[width=0.45\textwidth]{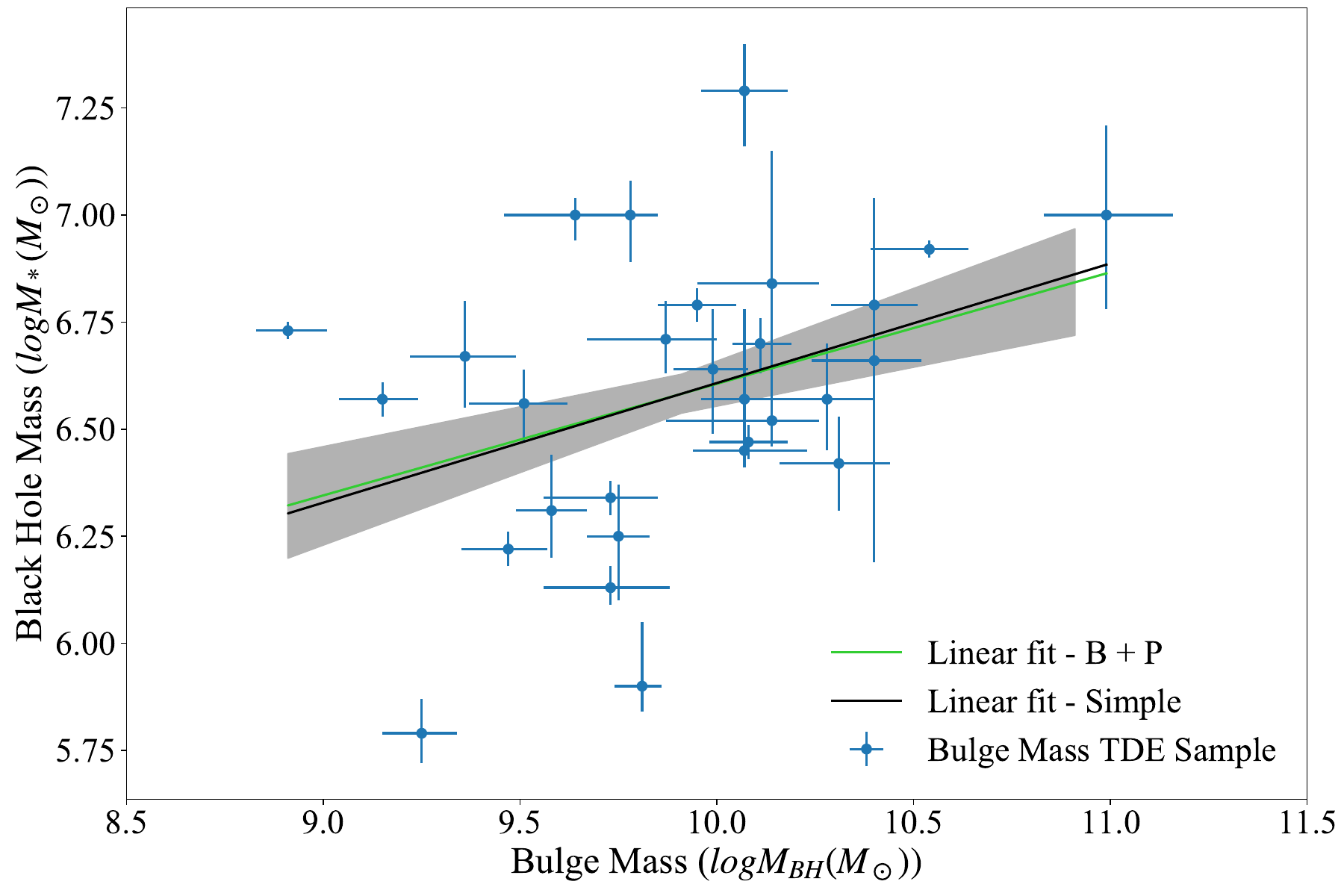}
\includegraphics[width=0.45\textwidth]{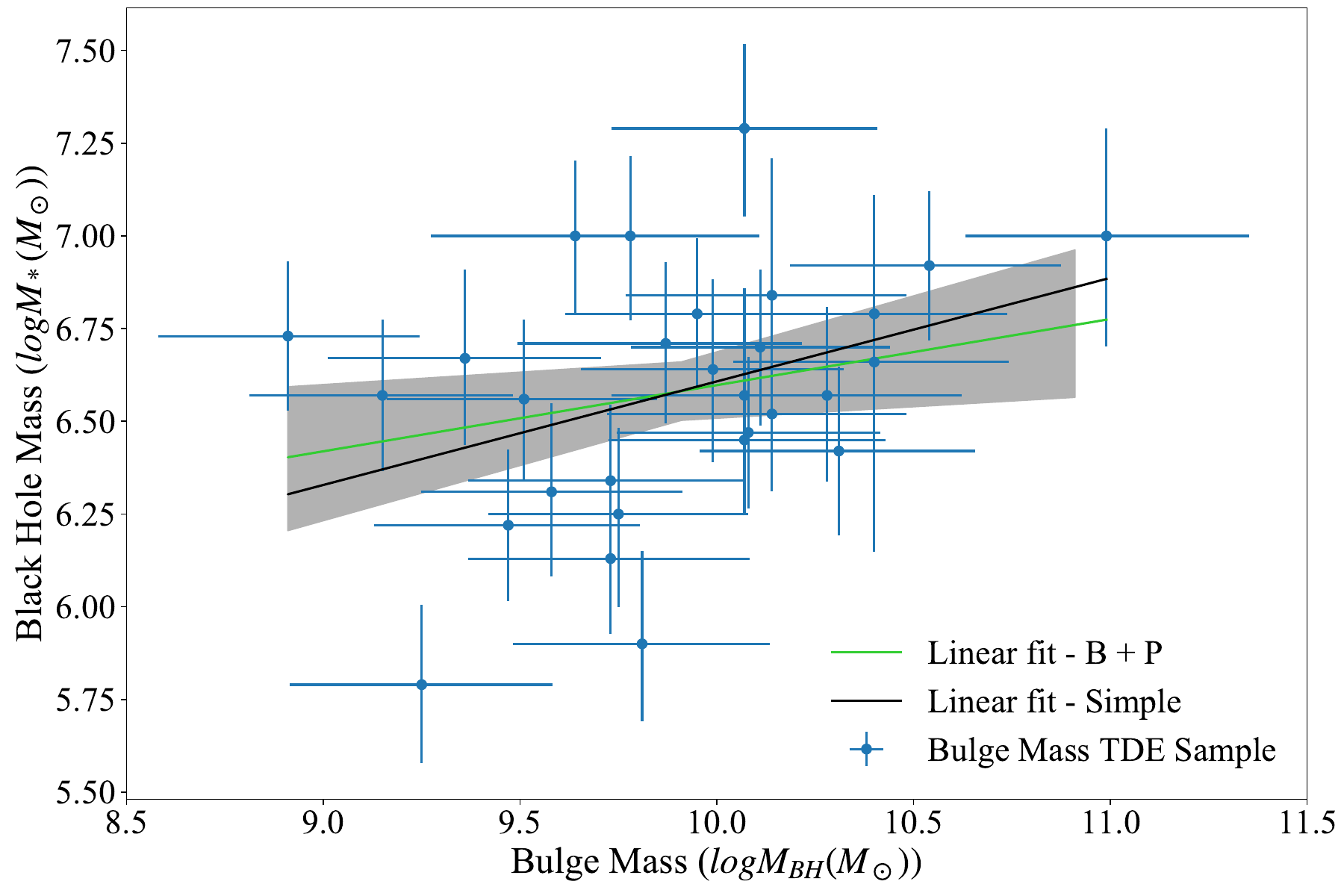}
\caption{SMBH mass as a function of host galaxy bulge mass, with a basic fit to the correlation in black. The green line and grey region represent the median and 90\% confidence interval calculated through Bootstrapping and Perturbation (P + B) of the data. The top panel includes only our derived statistical errors, while the bottom includes estimates of the systematic error on each quantity from \textsc{mosfit} and \textsc{prospector}. The black line corresponds to equation \ref{eq:tde-simple}, while the green line in the lower panel corresponds to equation \ref{eq:tde-error}.}
\label{f:bulgecomp}
\end{figure}

SMBH mass as a function of host galaxy bulge mass for the TDE sample is illustrated in Figure \ref{f:bulgecomp}. With no resampling, we find a Spearman coefficient $r_s = 0.35$ with a significance of $P_s = 0.06$. The best-fit straight line, which we call the `simple' fit, has a gradient of 0.28 and a y-intercept of 3.82. Expressed in the same form as \citet{kormendy}, this gives a relationship between SMBH mass and host galaxy bulge mass:
\begin{equation}\label{eq:tde-simple}
    \frac{M_{\rm BH}}{10^9M_{\odot}} = 0.01 \left(\frac{M_{\rm Bulge}}{10^{11}M_{\odot}}\right)^{0.28} .
\end{equation}

On the same plot, we show the results after applying the MC resampling alongside bootstrapping with replacement, with the average fit shown as a green line. If we consider only the statistical errors in the data, we calculate a Spearman coefficient of $r_s = 0.33^{+0.09}_{-0.09}$ with a significance of $P_s = 0.13^{+0.10}_{-0.10}$. A correlation is usually considered significant only if $p<0.05$. From this fitting procedure we see that approximately 32$\%$ of trials were formally significant, suggestive of a possible positive correlation at moderate statistical significance for this sample. If we include also the systematic errors on each parameter \citep{leja, mockler}, we instead find, $r_s = 0.23^{+0.15}_{-0.15}$ with a significance of $P_s = 0.31^{+0.33}_{-0.27}$. When looking at the percentage of significant trials from the fitting procedure, after considering both statistical and systematic uncertainties, it remains that approximately 15\% of trials were significant (p $<$ 0.05) despite the limitations of the large systematic errors.

The bootstrapping + perturbation best fit, illustrated in Figure \ref{f:bulgecomp}, has a gradient of $0.26^{+0.07}_{-0.07}$ (without systematic errors), or $0.18^{+0.11}_{-0.11}$ if we include the estimated systematics. The most conservative relationship between SMBH mass and host galaxy bulge mass was therefore found to lie within the 1$\sigma$ range

\begin{equation}\label{eq:tde-error}
    \frac{M_{\rm BH}}{10^9M_{\odot}} = (0.01^{+0.01}_{-0.02}) \cdot \left(\frac{M_{\rm Bulge}}{10^{11}M_{\odot}}\right)^{(0.18 \pm 0.11)}.
\end{equation}

In order to determine how well BH masses can be derived using the relationship, the intrinsic scatter of the data around this line was calculated. We used the method of \cite{mcconnell} (their equation 4), taking into account the statistical and systematic errors in each parameter, and varying the intrinsic scatter term for each fit until $\chi^2=1$ was achieved. We find that the intrinsic scatter of the data around equation \ref{eq:tde-error} is 0.22\,dex. If we instead use the slope from equation \ref{eq:tde-simple}, we find a scatter of 0.20\,dex. In summary,  these equations can be used to estimate the SMBH masses of TDE hosts from their bulge masses to within around 0.2\,dex.

\subsection{TDE Sample in Combination with High-mass Regime}
\label{s:combsamp}

\begin{figure*}
\centering
\includegraphics[width=\textwidth]{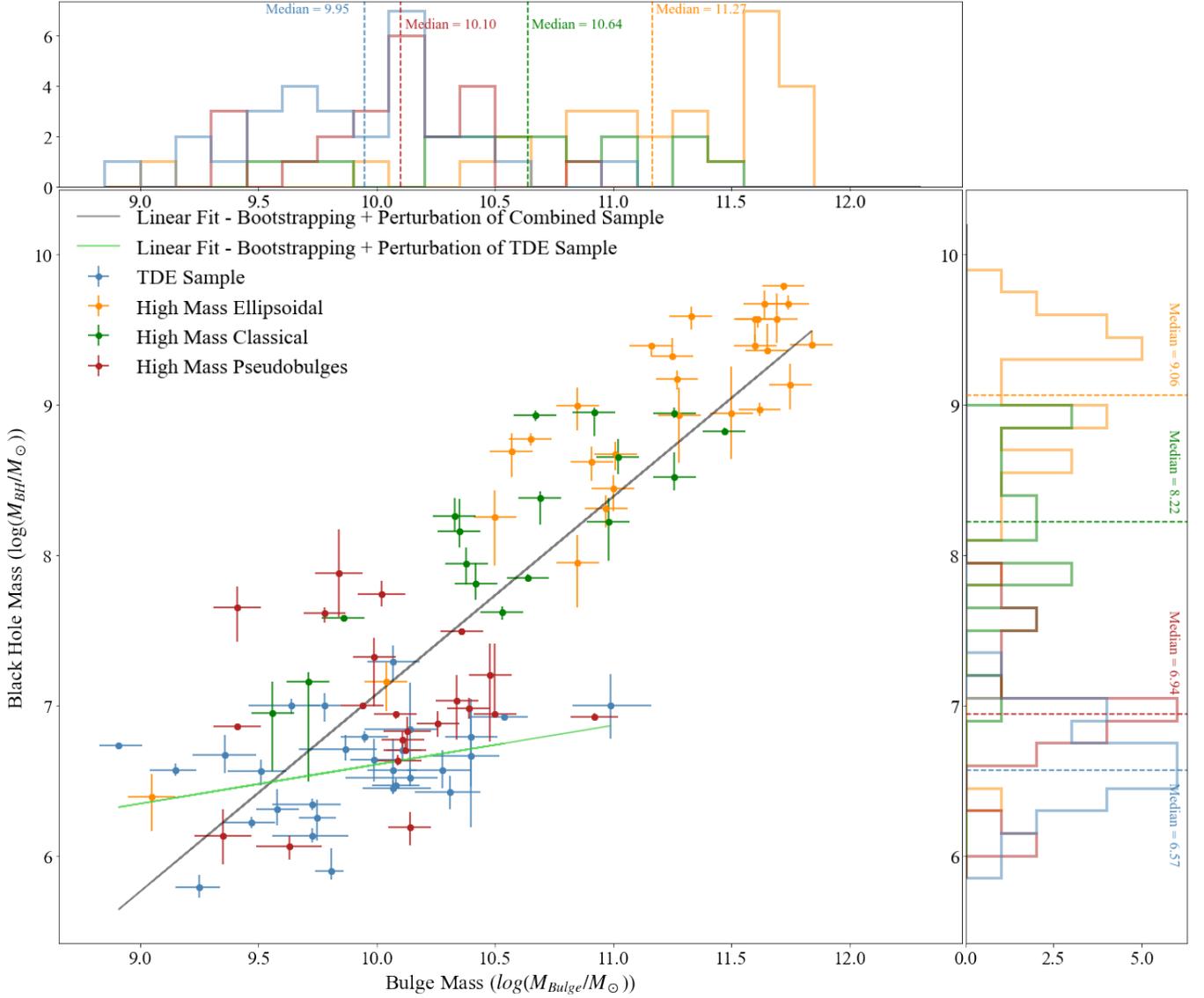}
\caption{\small SMBH mass as a function of host galaxy bulge mass for the TDE sample (blue, showing statistical errors only) and the high-mass regime: ellipsoidal galaxies (orange), classical bulges (green) and pseudobulges (red). The black line (equation \ref{eq:combined}) shows the averaged fit to the combined sample, after bootstrapping and MC resampling, while the green line shows the equivalent quantity for the TDEs only (equation \ref{eq:tde-error}). Histograms of the bulge and SMBH mass distributions are shown on their respective axes, where the median of each distribution is indicated via a dashed line.}
\label{big_scatter}
\end{figure*}

In order to determine where TDEs and their hosts fall within the mass distributions of other SMBH samples, and whether they can improve the calibration of the SMBH--stellar mass relationship at the low end of the mass spectrum, the TDE sample was combined with that of the high-mass regime. We used the data from \cite{kormendy}, which is sourced from an inventory of host galaxies with SMBH measurements based on stellar dynamics, maser disk dynamics, ionized gas dynamics, or CO molecular gas disk dynamics. The inventory was categorised into morphological types: elliptical galaxies, classical bulges and pseudobulges. The data in total comprise 67 sources. We omitted any data that \cite{kormendy} neglected to fit, and any sources that they identified as ongoing mergers. 

Figure \ref{big_scatter} shows SMBH mass as a function of host galaxy bulge mass for the combined sample. TDE host galaxies have the lowest SMBH and bulge masses, and populate the previously sparsely-filled region of the plot with $M_{\rm bulge}\lesssim10^{10}$\,M$_\odot$ and $M_{\rm BH}\lesssim10^7$\,M$_\odot$.
We apply the same statistical tests to this sample as in the previous section.
Unsurprisingly, given the known correlation in the high-mass sample, the combined sample is found to have a strong correlation with $p\ll0.05$.

The gradient derived from the MC sampling and bootstrapping, illustrated in Figure \ref{big_scatter} but including systematic errors, is $1.31^{+0.03}_{-0.03}$, with a y-intercept of $-6.06^{+0.31}_{-0.31}$. The best fit relationship across the whole sample was therefore found to be

\begin{equation}\label{eq:combined}
    \frac{M_{\rm BH}}{10^9M_{\odot}} = (0.22^{+0.27}_{-0.26}) \cdot \left(\frac{M_{\rm Bulge}}{10^{11}M_{\odot}}\right)^{(1.24 \pm0.05)}
\end{equation}

\section{Discussion}
\label{s:discussion}

Compared to other methods of measuring BH mass, we find that TDEs can be used to directly target low mass SMBHs. The median SMBH mass found for our TDE sample is $10^{6.57}\,{\rm M}_{\odot}$ \citep{nicholl2022} compared to the median of the high mass regime $10^{8.07}\,{\rm M}_{\odot}$. This supports the idea that observations of lower mass SMBHs ($\sim 10^6\,{\rm M}_{\odot}$) are more common around TDEs, due to the critical mass resulting from the the tidal and gravitational radius inequality, $R_T > R_S$ \citep{hills}. Here we have demonstrated that these are also found in galaxies with low bulge masses, $\lesssim 10^{10.5}$\,$M_{\odot}$

The scaling of SMBH mass with host galaxy mass identified in our isolated TDE sample was statistically weaker than the very strong correlation found in the combined sample. Shown in Figure \ref{big_scatter}, the TDE sample visibly populates the lower half of the combined relationship, extending it to lower mass BHs. There is significant overlap between pseudobulges and our TDE sample, with the median pseudobulge SMBH mass laying much closer to that of the TDEs than the other components of the high mass regime. \cite{kormendy} removed the pseudobulge sample from their final relationship after concluding they did not satisfy the tight SMBH--host galaxy correlations that classical and elliptical bulges did.

For our purposes, the pseudobulge sample are interesting to consider due to their greater overlap in SMBH mass with the TDE hosts \citep{kormendy}. As a general definition, pseuodobulges are bulges which have morphologies reminiscent of disk galaxies -- as opposed to classical bulges, which appear similar to elliptical galaxies \citep{fisher, kormendy}. When fitted with a Sérsic profile, pseudobulges have a lower index, $n_b$, indicating a smaller bulge \citep{fisher} which, considering co-evolution, suggests a lower mass SMBH. Furthermore, \cite{hu} found that pseudobulges have a tight SMBH--velocity dispersion relation, but with a slope that is distinct from that measured in the classical bulges and elliptical galaxies. This suggests that SMBH growth in pseudobulges is slower than that in the early-type bulges \citep{hu, kormendy, jiang}. While they have been previously omitted from the SMBH--stellar mass relationship, with the addition of the TDE sample, it is clear that pseudobulges do fit within the combined sample. This further indicates the successful role TDEs play in extending the relationship to the lower mass end of the SMBH spectrum.

Comparing the final extended SMBH - stellar mass relationship derived for the combined sample, equation \ref{eq:combined}, to that derived in \cite{kormendy}, 
\begin{equation}
    \frac{M_{\rm BH}}{10^9 M_{\odot}} = (0.49^{+0.06}_{-0.05})\left(\frac{M_{\rm Bulge}}{10^{11} M_{\odot}}\right)^{1.16 \pm 0.08}
\end{equation}
we see that the normalisation is in $\sim 1 \sigma$ agreement, but the difference in the exponent is significant. Moreover, the exponent we derived is $\sim 5 \sigma$ away from being linear, compared to the $\sim 2 \sigma$ difference in \cite{kormendy}. Therefore, with the combined sample we find a steeper SMBH--stellar mass relationship, indicative of a faster scaling SMBH mass. It is worth noting that while \cite{kormendy} omitted pseudobulges in their final relationship due to finding a reduced correlation between SMBHs, when removing pseudobulges from our combined sample there are no significant changes in the statistical values shown in Table \ref{t:stats}. 

Although we have extended the mass range, we are not claiming that our measured relationship is more reliable than the widely used relation from \citet{kormendy}. In particular, the faster scaling in our sample relies on the accuracy of the SMBH mass measurements from \citet{nicholl2022}, and hence are limited by the approximations used in light curve modelling. However, the indicative results from our study suggest that if we can obtain accurate SMBH masses for future large TDE samples, these will have significant statistical power in constraining the slope of this fundamental relationship.

The comparative flatness between the TDE-only and combined samples could be due to a restrictive and small sample size for the TDEs. The TDE sample obtained is a good representation of all observed TDEs, but covers narrow ranges of SMBH ($10^{5.80} - 10^{7.37}\,{\rm M}_{\odot}$) and host galaxy bulge masses ($10^{8.91} - 10^{10.99}\,{\rm M}_{\odot}$). Recovering a correlation with substantial intrinsic scatter is difficult with only a small dynamic range. However, it is also possible that this shallower correlation is a true physical effect of the TDE sample; it could be caused by the bias TDEs have towards lower mass BHs due to the limiting Hills mass \citep{hills}. In addition, it can be seen in Figure \ref{big_scatter}, between a fixed bulge mass of $10^{10}\,{\rm M}_{\odot} - 10^{11}\,{\rm M}_{\odot}$, that the TDE SMBH masses are systematically lower than those of the elliptical and classical samples, by $\sim$ 1 dex. This further highlights the bias that TDEs have towards lower mass SMBHs. A larger sample is required to further understand the difference in relation in this crucial mass range. Furthermore, the comparative flatness could also be a result of the SMBH--bulge mass relationship changing at lower masses. For example, \cite{jiang} studied the scaling between host galaxy bulge luminosity and the mass of the central BH. Similar to our study, they found that the relation flattened in the lower mass region, in which the host bulges spanned a wide range of luminosities at a fixed BH mass. This is similar to what we see here: a wide range of bulge masses ($\sim$ $10^{9} - 10^{11}) M_{\odot}$ compared to a narrow band of BH masses ($\sim$ $10^{5.5} - 10^{7.5} M_{\odot}$). It is worth noting that additional studies also find evidence that bulges $\sim 10^{10} M_{\odot}$ contain black hole masses systematically lower than expected \citep{graham_dis,savorgnan}.

It is predicted that TDE observations will be revolutionised by upcoming surveys; the current detection rate of 10 year$^{-1}$ is set to increase to thousands year$^{-1}$ \citep{bricman, khabibullin}. Covering $\sim$ 10,000 deg$^2$ of sky each night, the upcoming LSST will survey six optical bands over 10 years, visiting each field approximately 900 times. Using the LSST simulation framework, \cite{bricman} predicted that LSST will discover between 35,000 - 80,000 TDEs during the survey. A large fraction will not have sufficient follow-up data for detailed individual study. However, the host galaxy information provided by LSST will be excellent, even with no targeted follow-up, due to the unprecedented survey depth. Hence, utilising the derived relationship between TDE host galaxy bulge mass and BH mass will allow us to efficiently analyse the upcoming sample. Given the narrow mass range of TDE hosts and the offset in SMBH masses for hosts in the range $10^{10}-10^{11}$\,M$_\odot$, we suggest to use the TDE-only relation to estimate SMBH masses for these events. This is accurate to $\sim 0.2$\,dex which is smaller than the systematic offset compared to the \cite{kormendy} relation. In time, the larger TDE sample will also enable us to improve the calibration of the SMBH--stellar mass relationship for TDEs, using those LSST events which are sufficiently well-observed for independent BH mass estimates.

\section{Conclusions}

We have modelled photometric data of 29 TDE host galaxies, deriving the total stellar mass and bulge mass for each event. The SMBH--stellar mass relationship was investigated for both the TDE sample and the TDE sample combined with that of the high-mass regime from \cite{kormendy}. Our main conclusions are as follows:

\begin{itemize}
    \item TDEs are successful in providing direct measurements of low mass SMBHs, with a median observed SMBH mass of $10^{6.57}\,{\rm M}_{\odot}$ compared to $10^{8.07}\,{\rm M}_{\odot}$ for previous SMBH samples.
    \item We find bulge masses in the range of $10^{8.91} - 10^{10.99} M_{\odot}$ with a median of $10^{9.95}\,{\rm M}_{\odot}$.
    \item We find a correlation between TDE host SMBH and bulge masses, with moderate statistical significance. The simple best-fit relation has a power-law slope of 0.28. After applying statistical tests and considering systematic errors, we show that the slope lies in the range $0.18 \pm 0.11$ ($1\sigma$) and the relation has an intrinsic scatter of 0.2 dex. 
    \item The relative flatness of this relation is different from that seen at higher SMBH mass, possibly reflecting event horizon suppression of the TDE rate.
    \item Combining the TDE sample with higher mass SMBHs, the strong positive correlation between bulge and SMBH mass is unsurprisingly recovered. However, TDEs prove to be highly constraining of the slope at the low mass end, and our sample shows possible evidence for a faster scaling of SMBH mass with bulge mass than previous studies, warranting further investigation with future large samples.
\end{itemize}

Our results indicate that TDEs are promising future probes for low mass SMBHs. With the projected detection rate of thousands per year, TDEs present an opportunity to explore a population of low-mass SMBHs that may otherwise lay dormant in the centre of their galaxies. The sample of TDEs and their hosts will continue to grow, enabling further understanding of the relationships and co-evolution between them.

\section*{Acknowledgements}

We thank an anonymous referee for their comments that improved the manuscript, and Joel Leja for helpful discussions about systematic uncertainties in the stellar population systhesis.
MN is supported by funding from the European Research Council (ERC) under the European Union’s Horizon 2020 research and innovation programme (grant agreement No.~948381) and by a Fellowship from the Alan Turing Institute.

%%%%%%%%%%%%%%%%%%%%%%%%%%%%%%%%%%%%%%%%%%%%%%%%%%
\section*{Data Availability}

This article is based on existing publicly available data.



\bibliographystyle{mnras}
\bibliography{refs} % if your bibtex file is called example.bib

%\begin{thebibliography}{99}
%\end{thebibliography}

%%%%%%%%%%%%%%%%%%%%%%%%%%%%%%%%%%%%%%%%%%%%%%%%%%

%%%%%%%%%%%%%%%%% APPENDICES %%%%%%%%%%%%%%%%%%%%%

\appendix

\section{Archival photometric data and fits}
\label{a:A}

\begin{figure*}
\centering
	\includegraphics[width=0.33\textwidth]{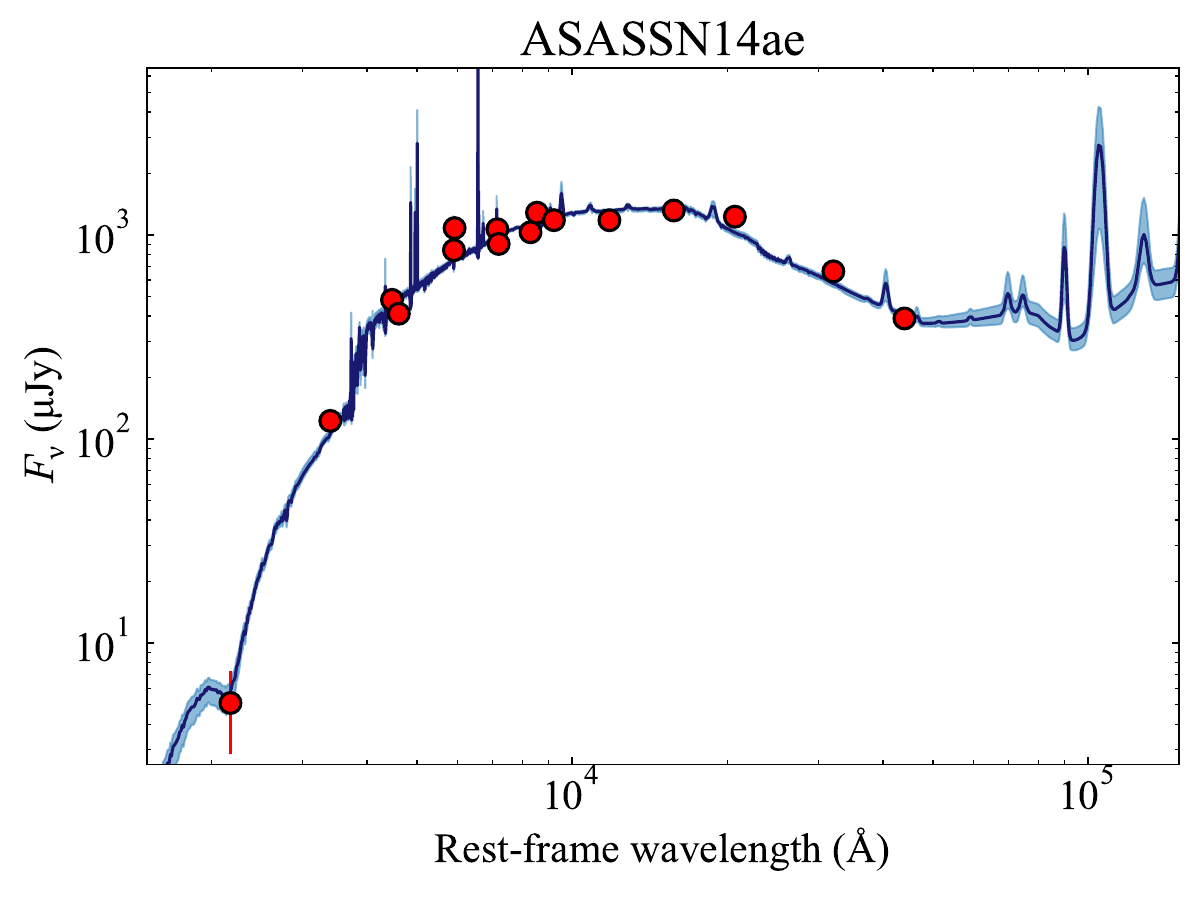}
	\includegraphics[width=0.33\textwidth]{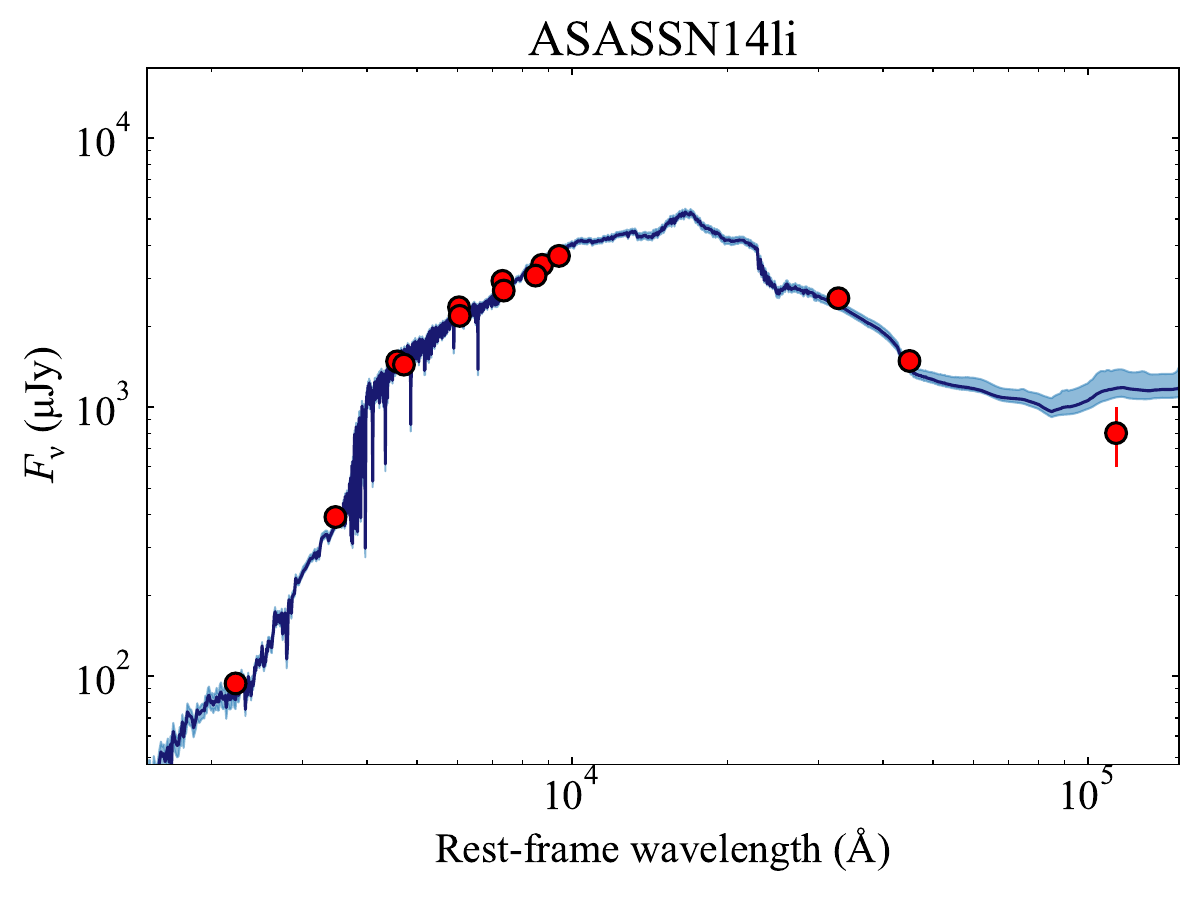}
	\includegraphics[width=0.33\textwidth]{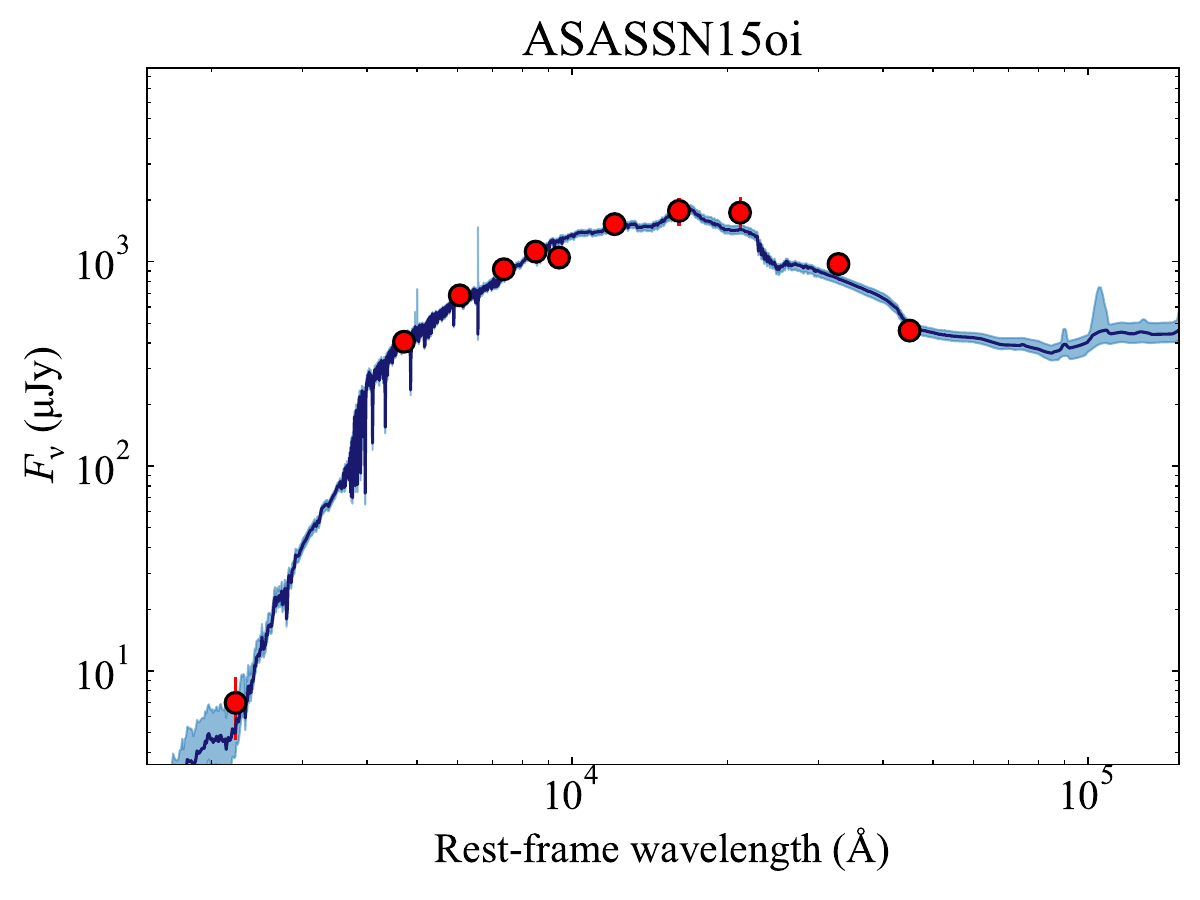}
	\includegraphics[width=0.33\textwidth]{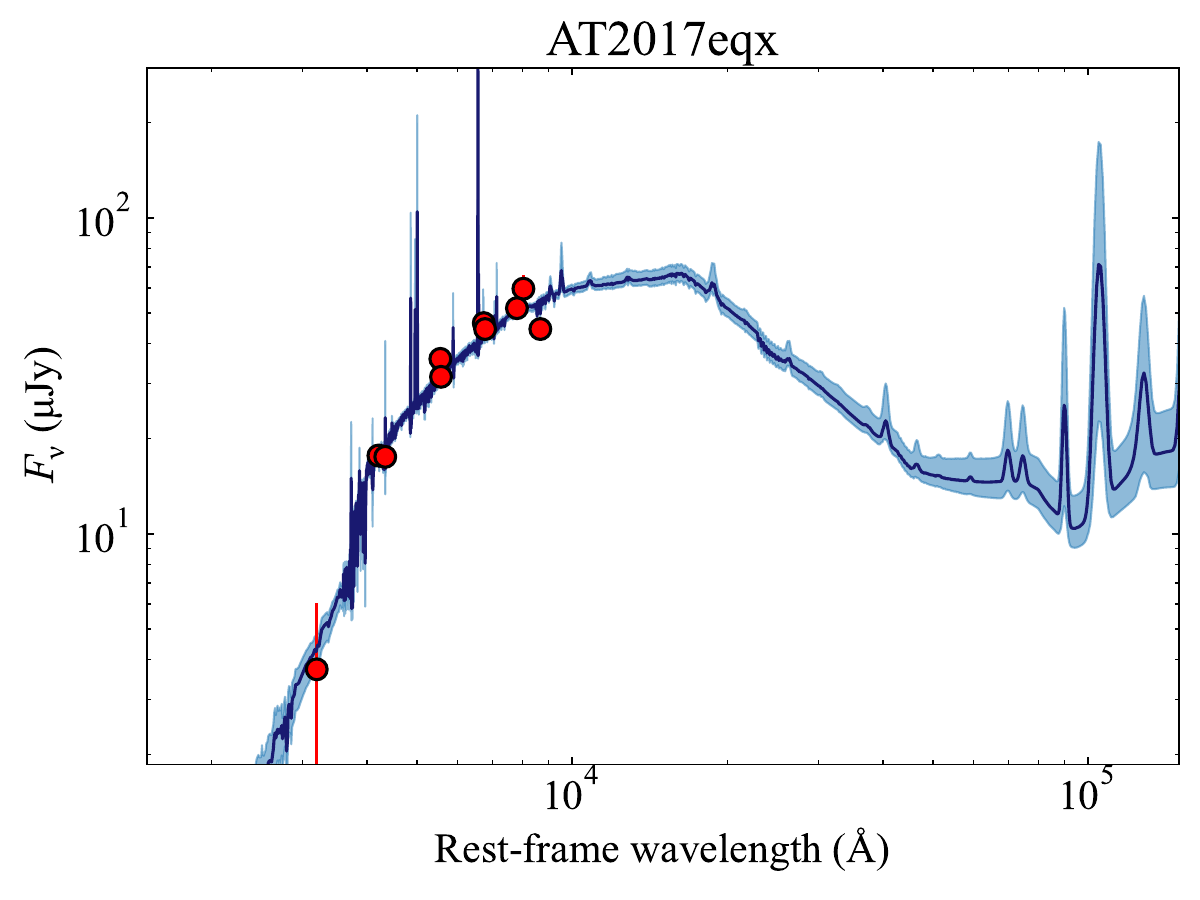}
	\includegraphics[width=0.33\textwidth]{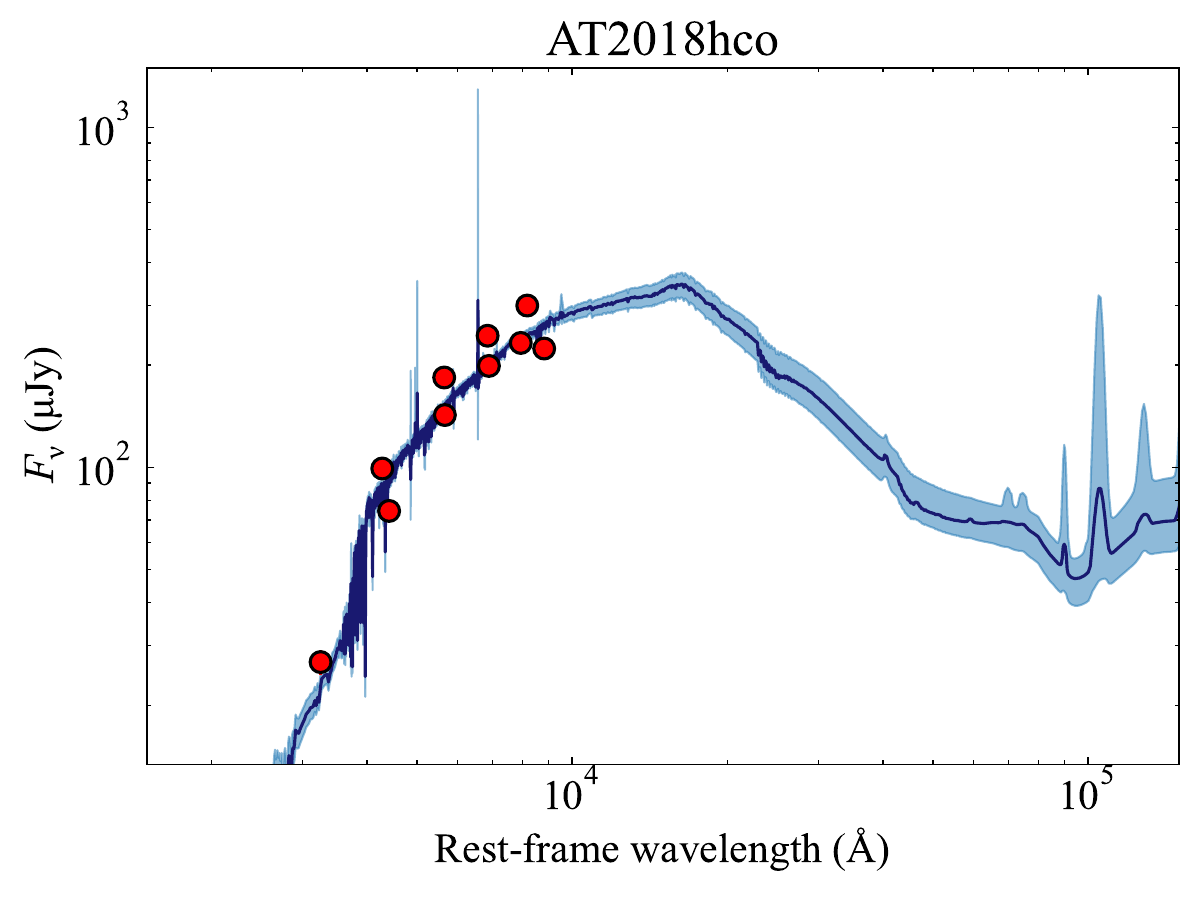}
	\includegraphics[width=0.33\textwidth]{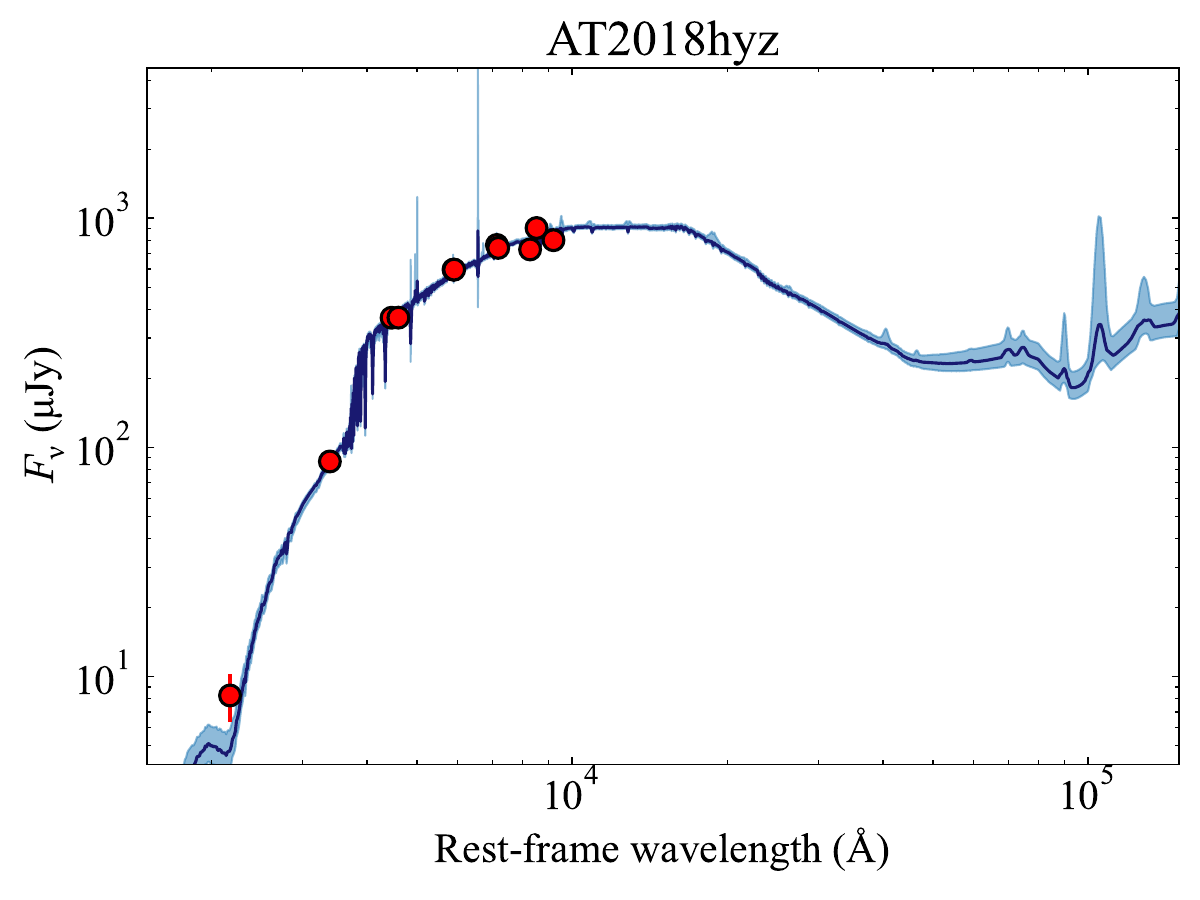}
	\includegraphics[width=0.33\textwidth]{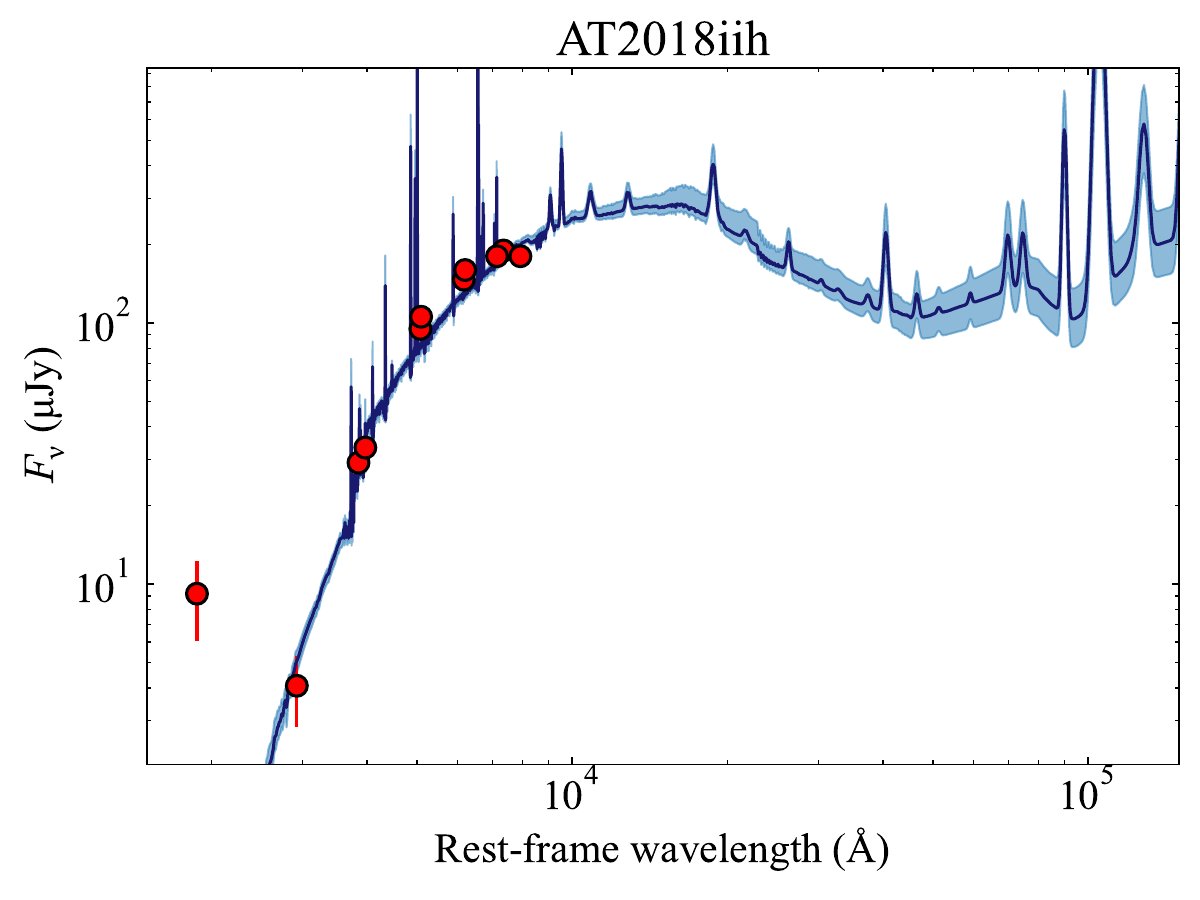}
	\includegraphics[width=0.33\textwidth]{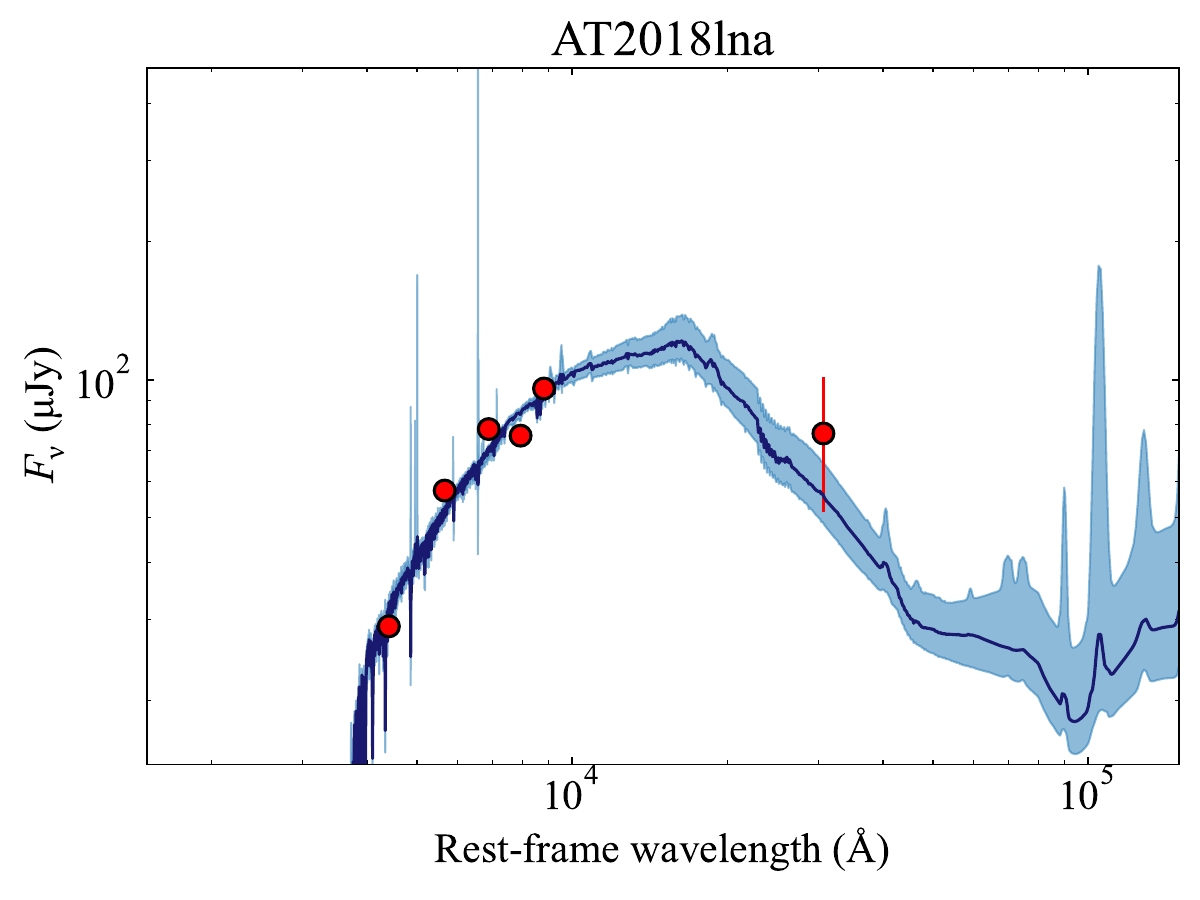}
	\includegraphics[width=0.33\textwidth]{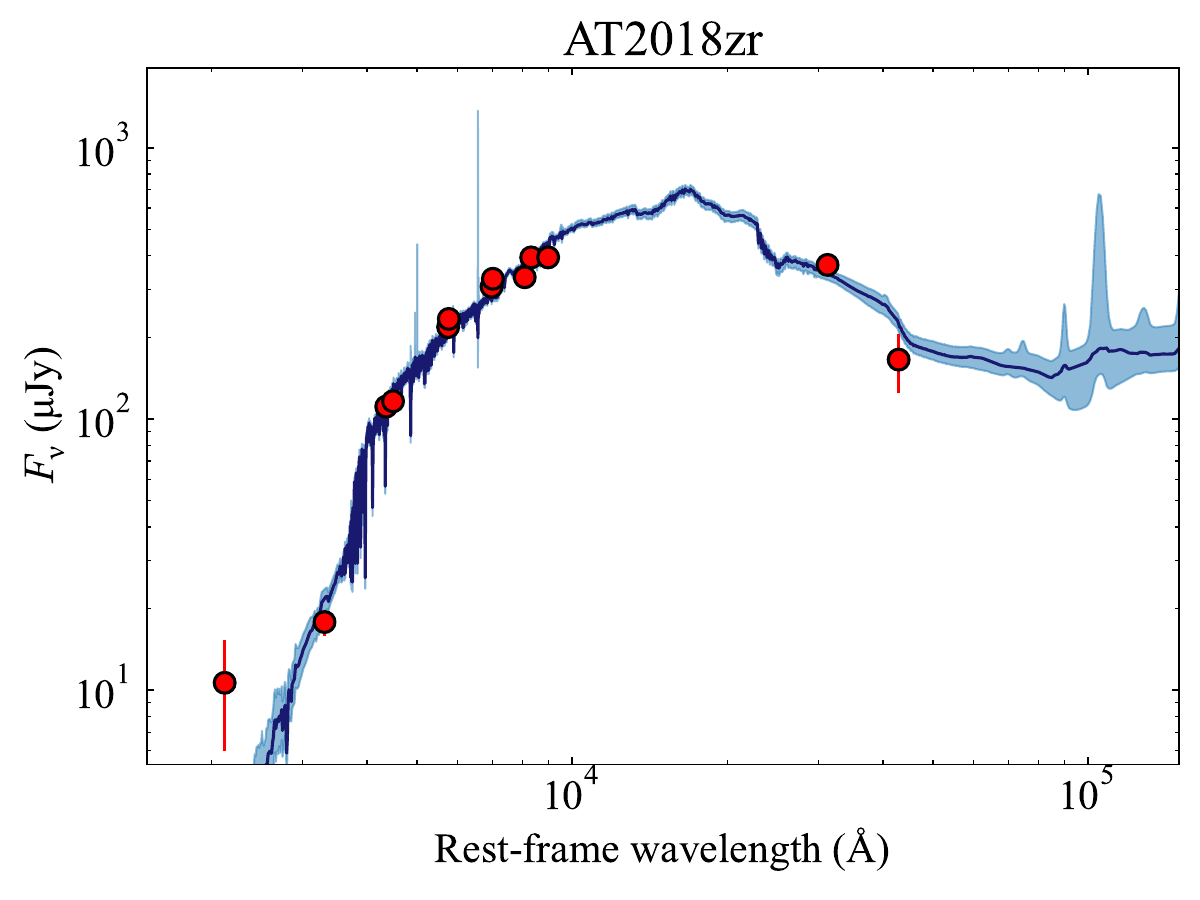}
	\includegraphics[width=0.33\textwidth]{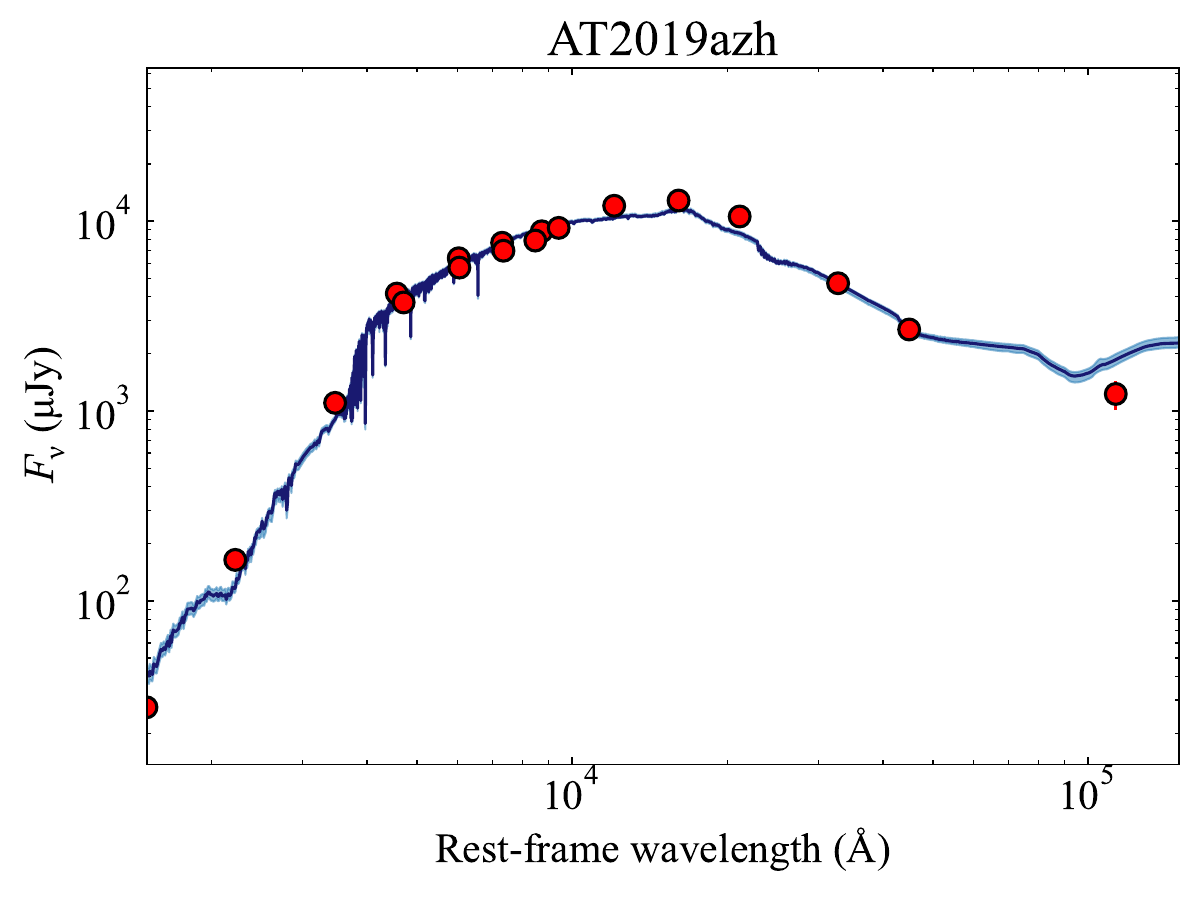}
	\includegraphics[width=0.33\textwidth]{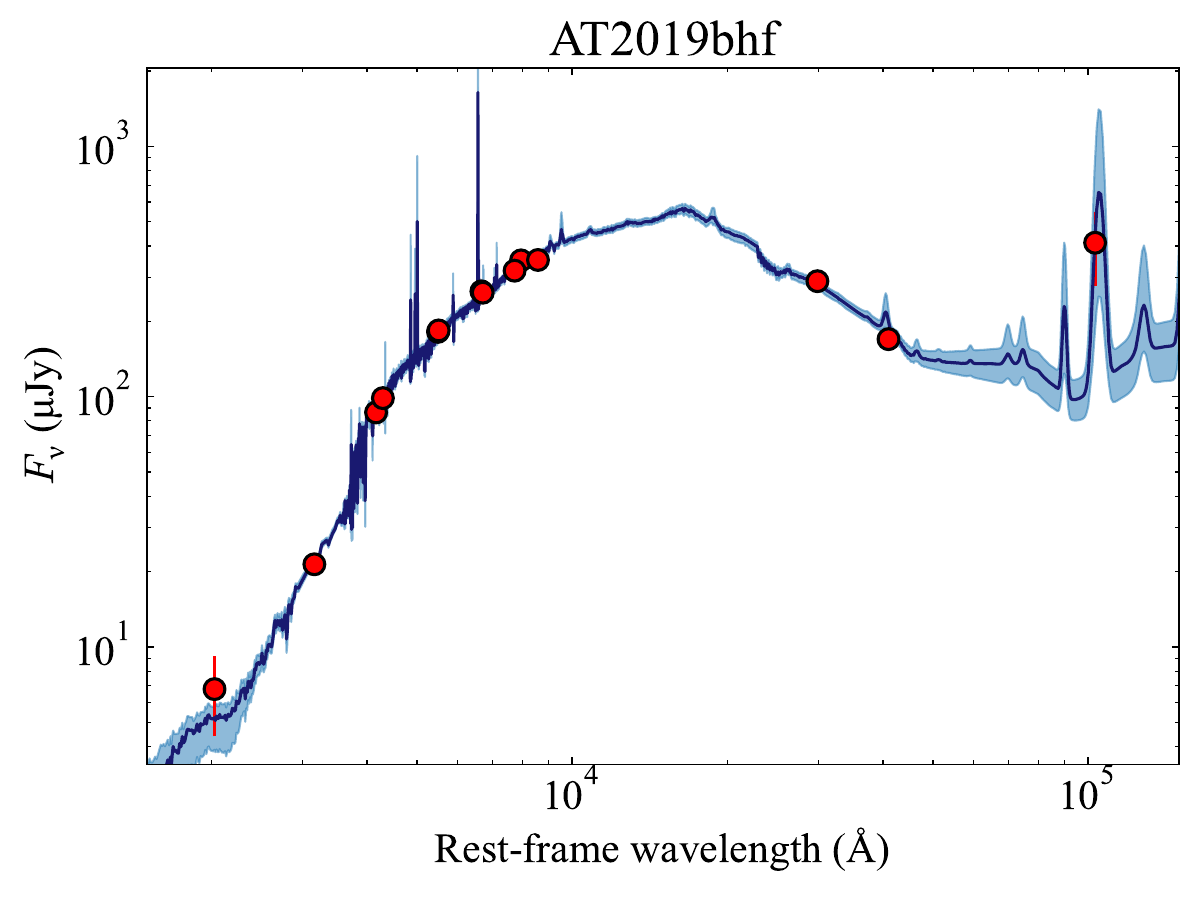}
	\includegraphics[width=0.33\textwidth]{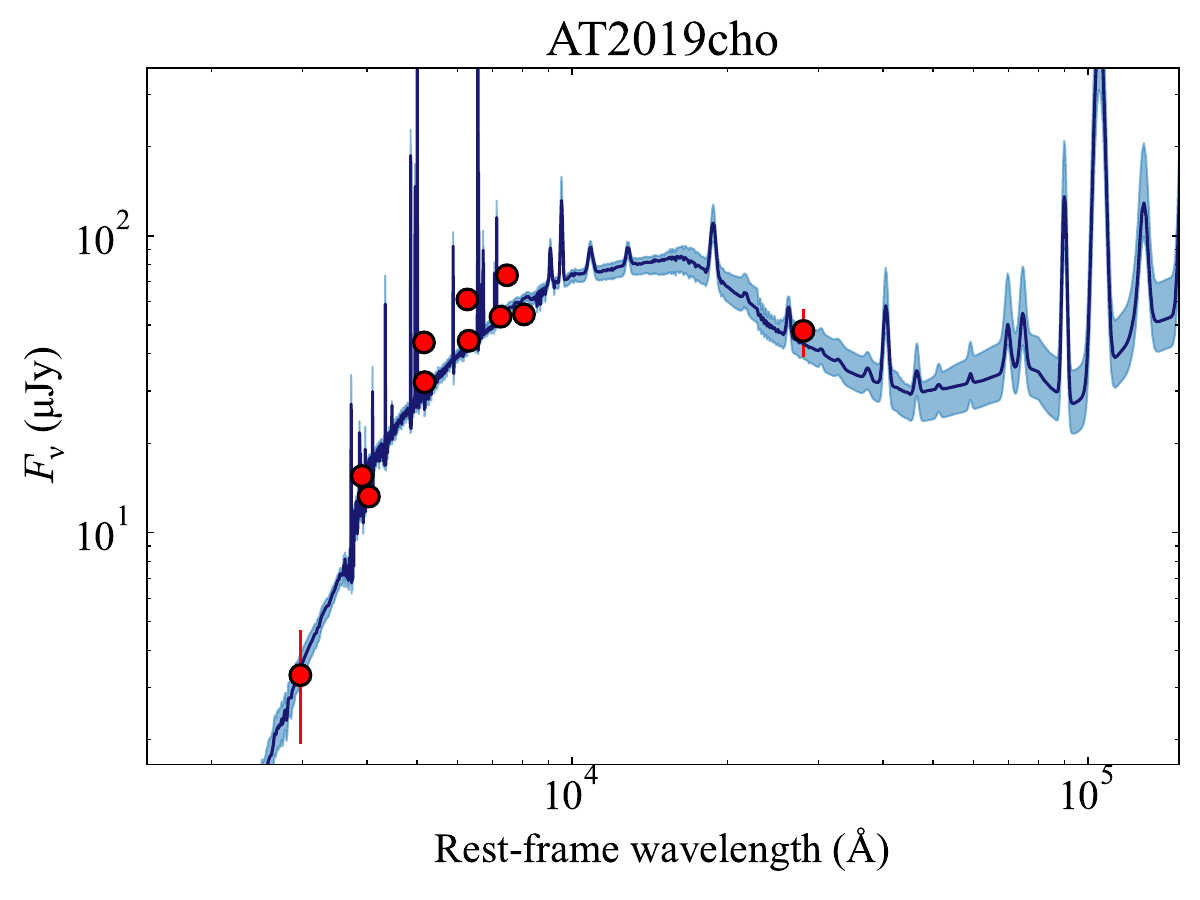}
	\includegraphics[width=0.33\textwidth]{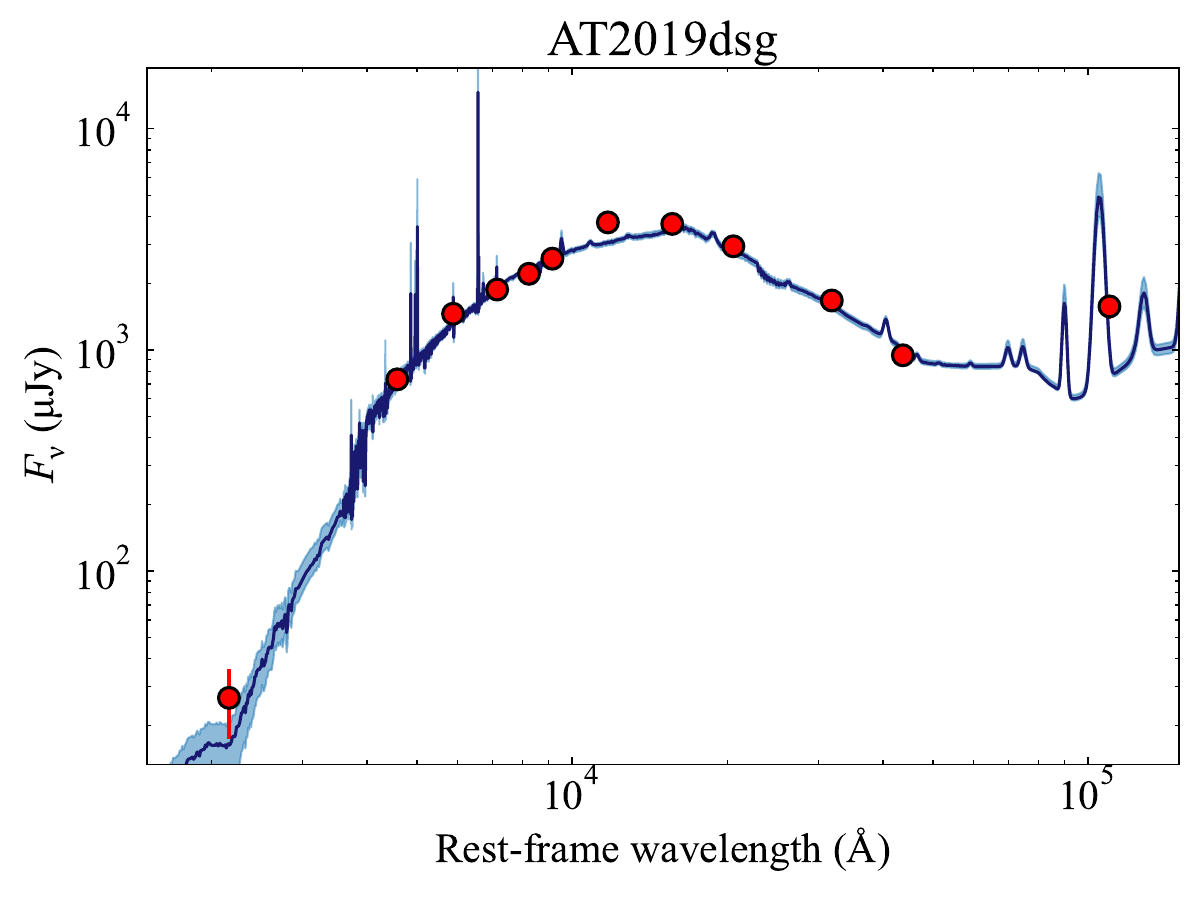}
	\includegraphics[width=0.33\textwidth]{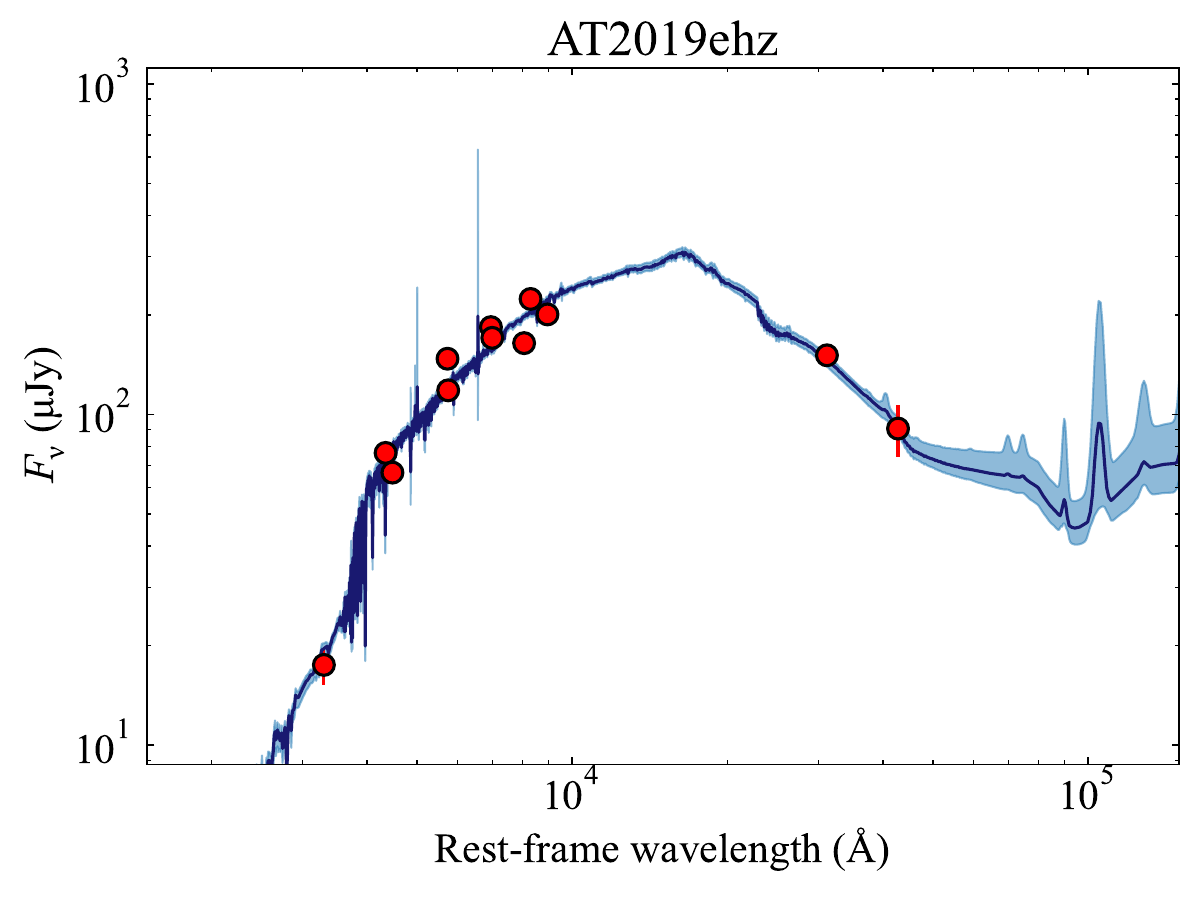}
	\includegraphics[width=0.33\textwidth]{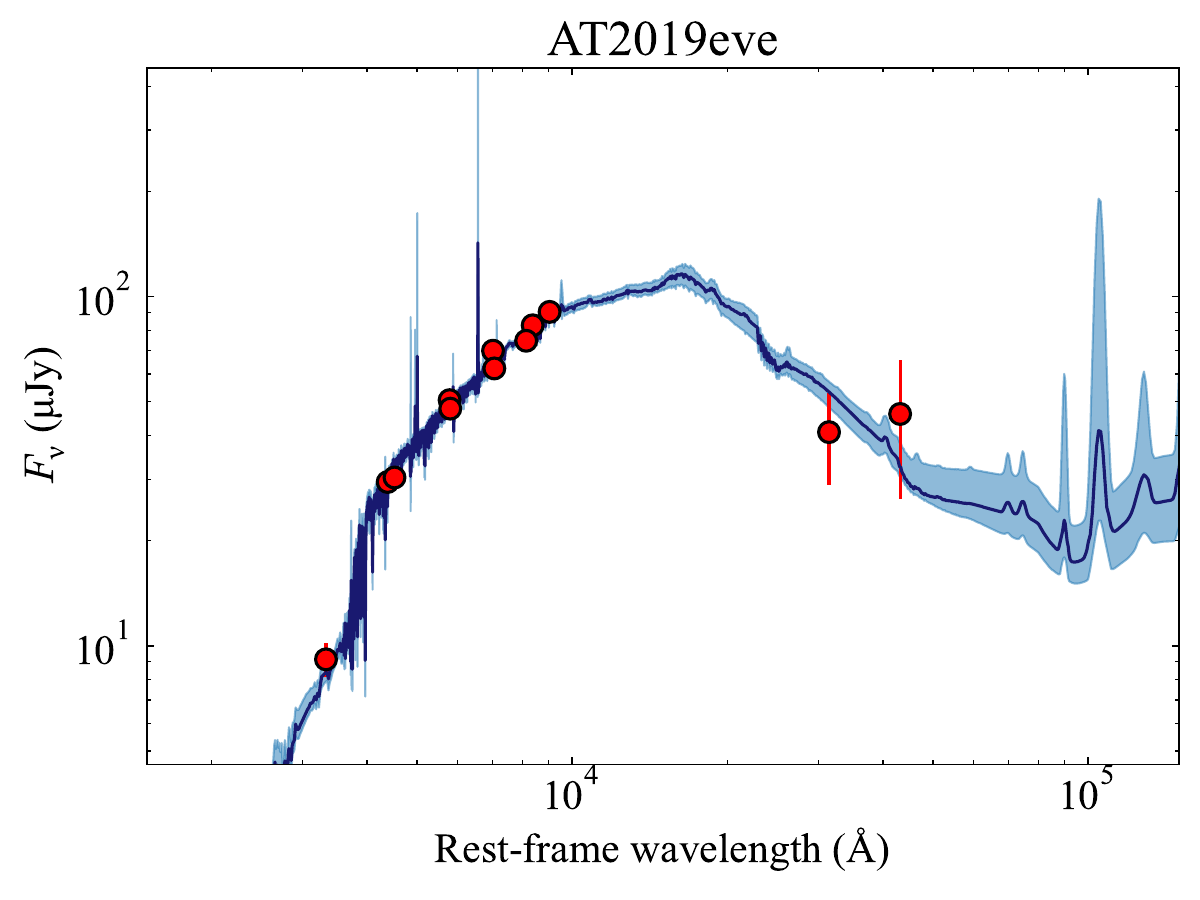}
	\caption{Archival photometry of each host galaxy (red) and the best fit model (dark blue) with the 16th to 84th percentiles of the distribution of model draws from the posterior (shaded region), derived through \textsc{Prospector}.}
    \label{fig:sed1}
\end{figure*}

\begin{figure*}
\centering
	\includegraphics[width=0.33\textwidth]{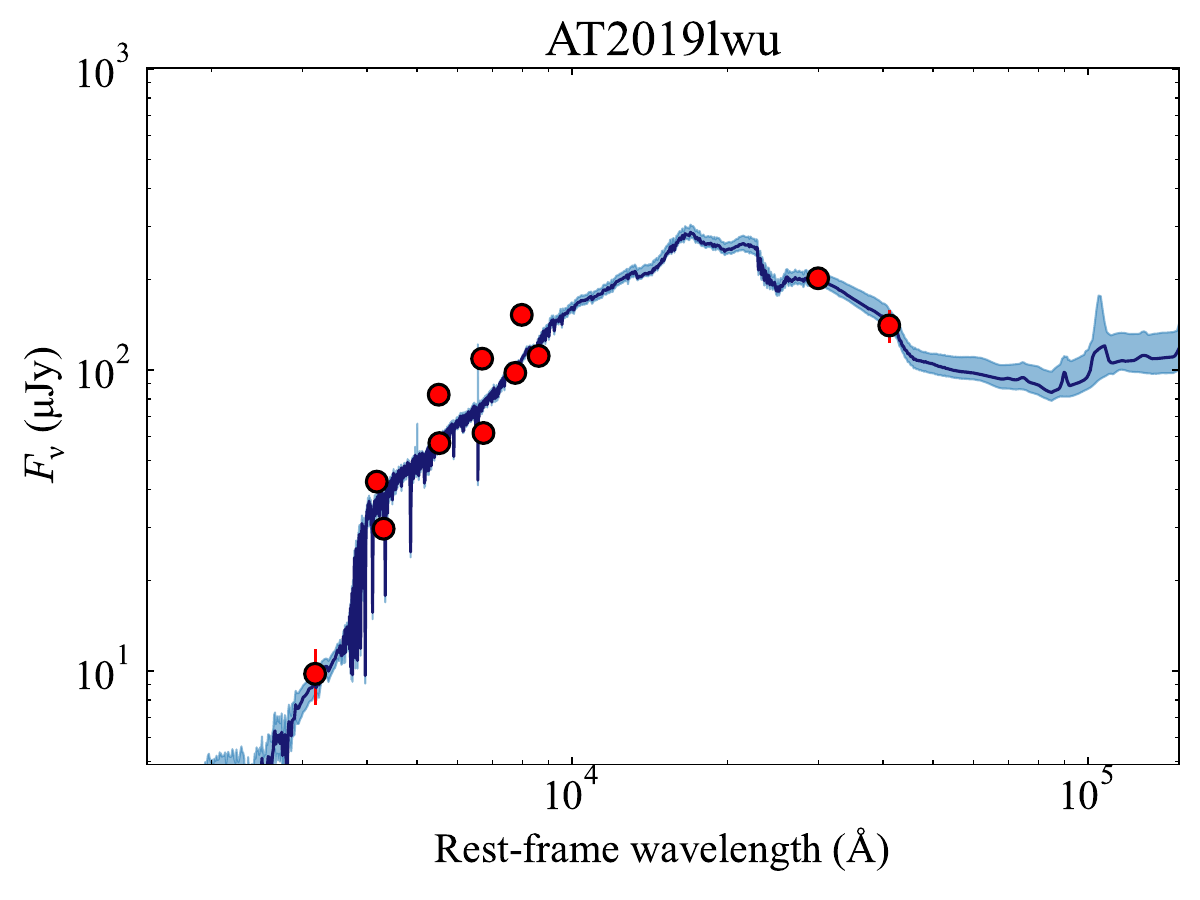}
	\includegraphics[width=0.33\textwidth]{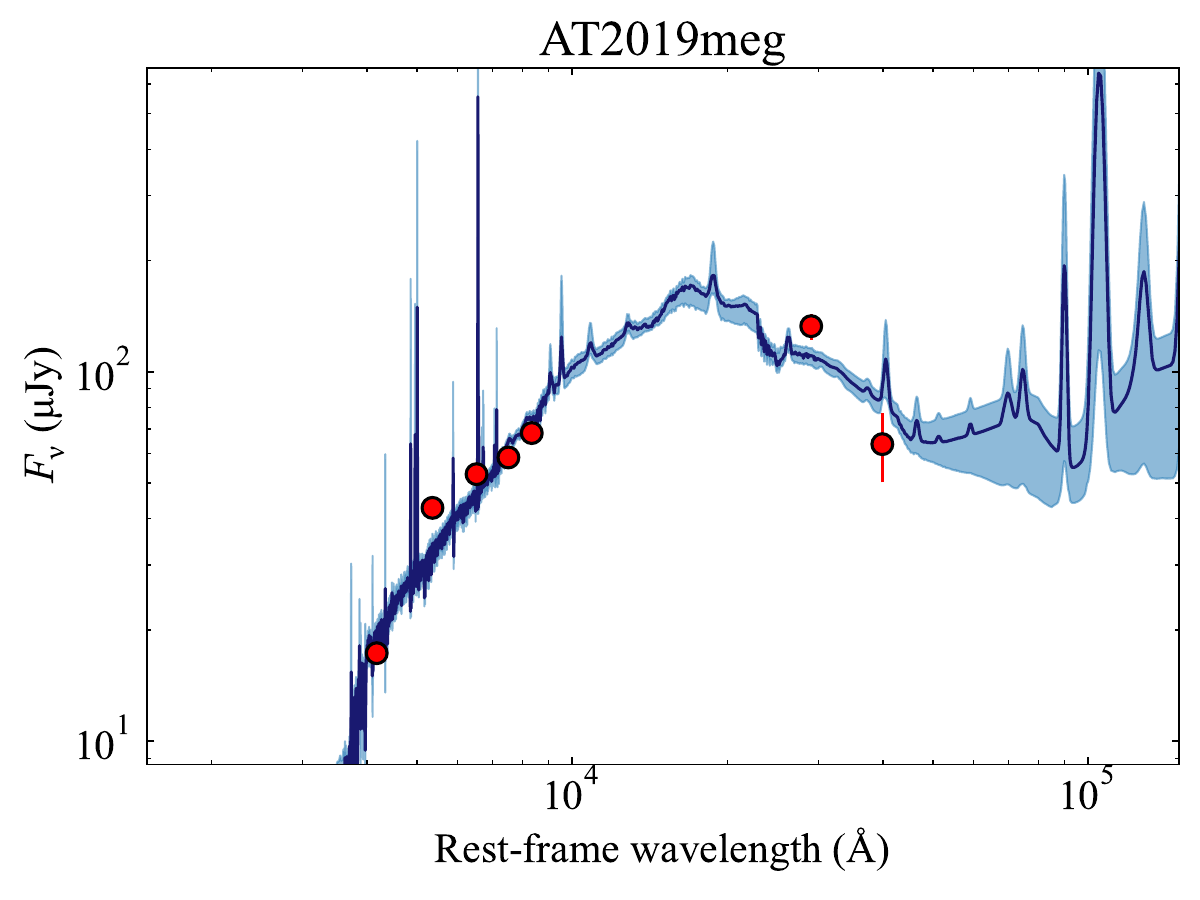}
	\includegraphics[width=0.33\textwidth]{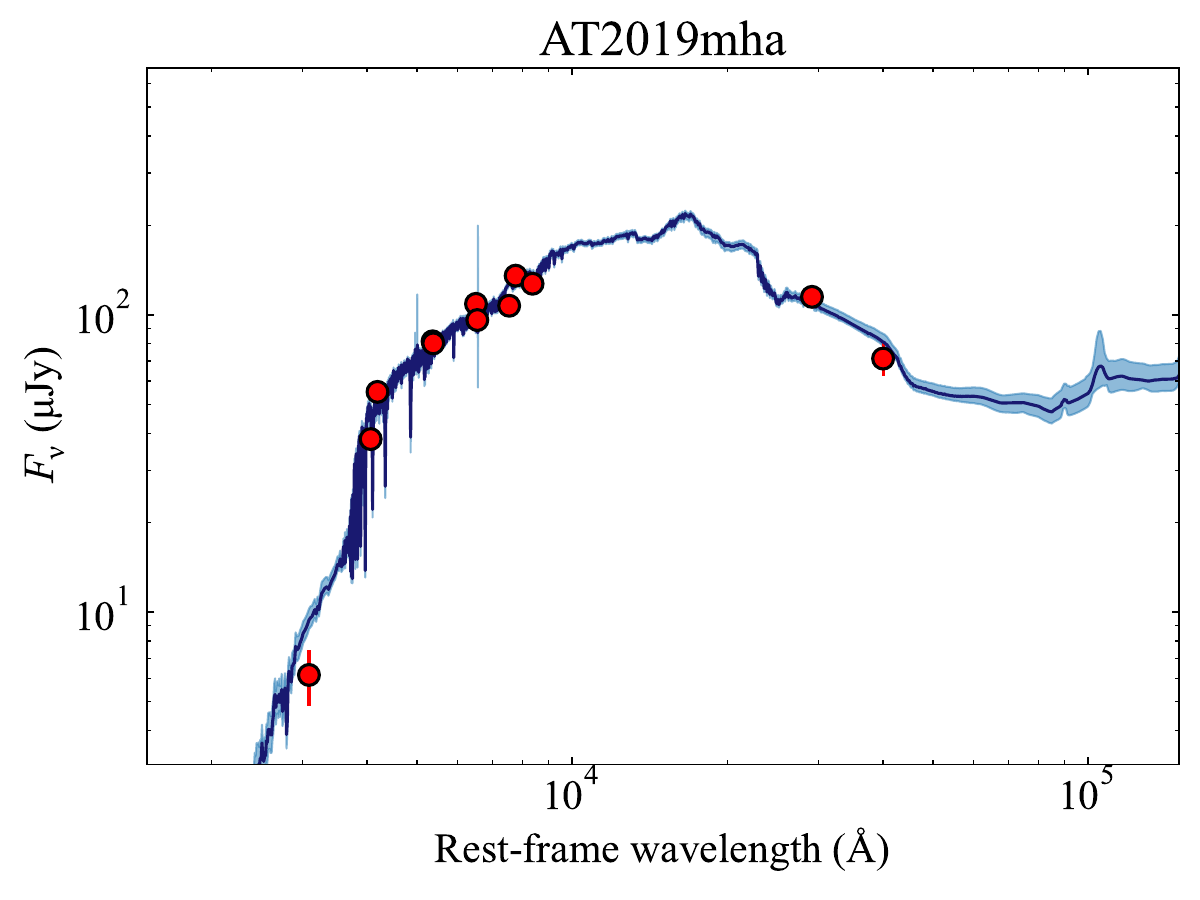}
	\includegraphics[width=0.33\textwidth]{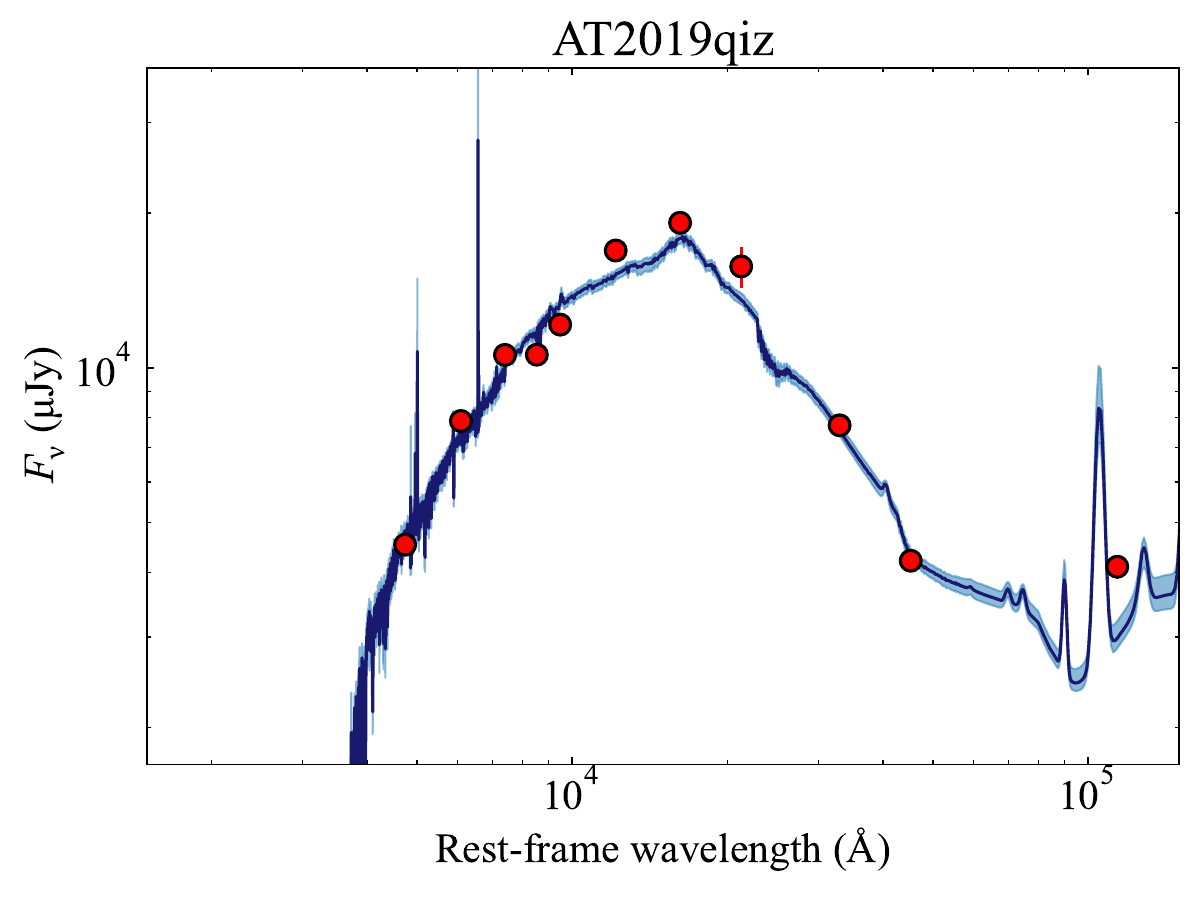}
	\includegraphics[width=0.33\textwidth]{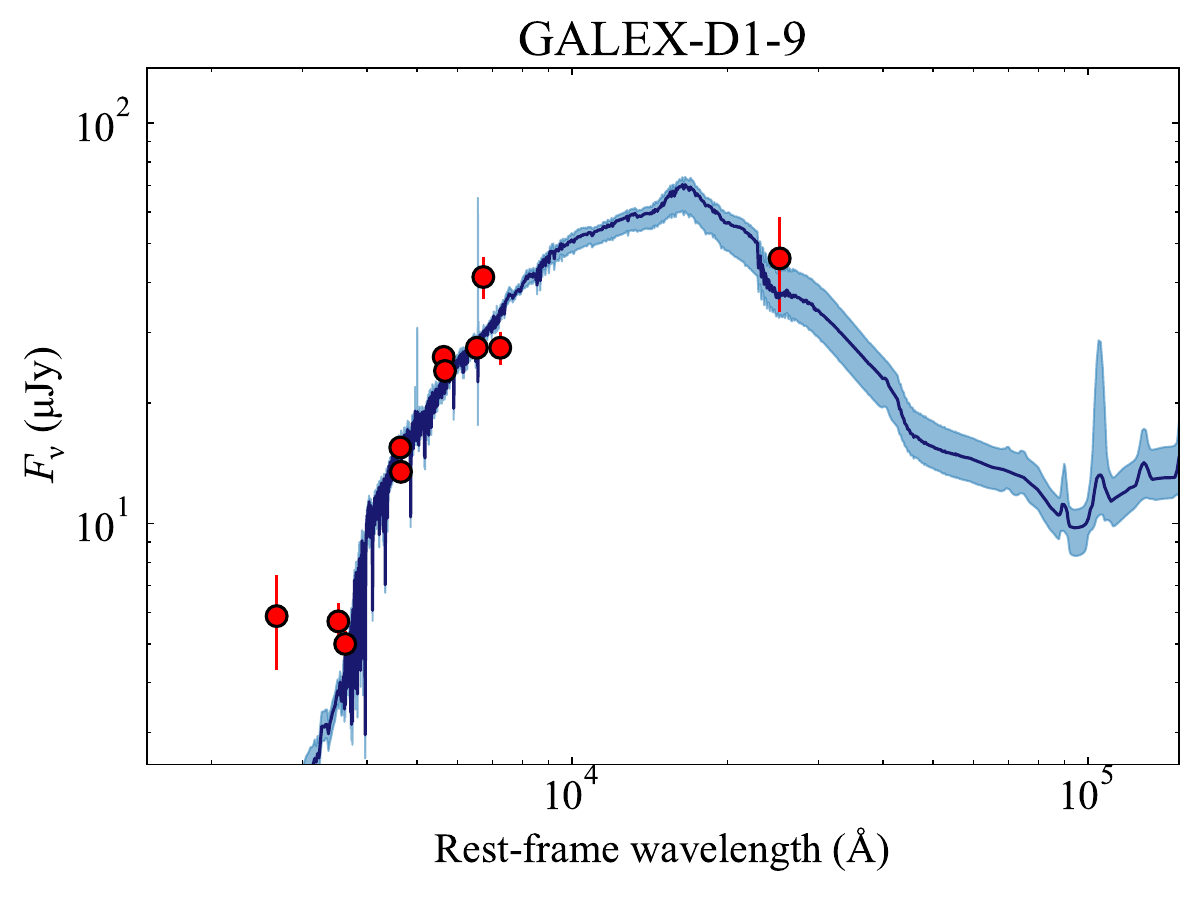}
	\includegraphics[width=0.33\textwidth]{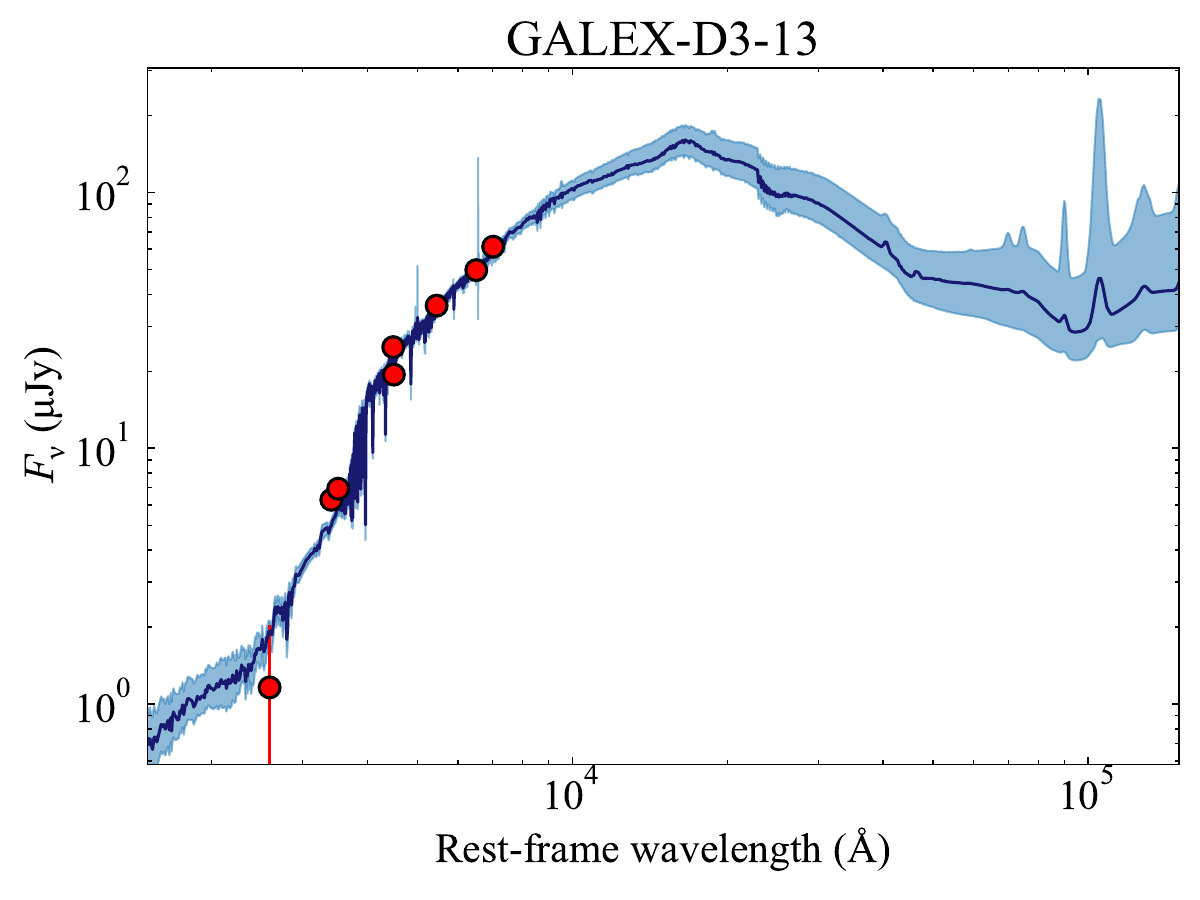}
	\includegraphics[width=0.33\textwidth]{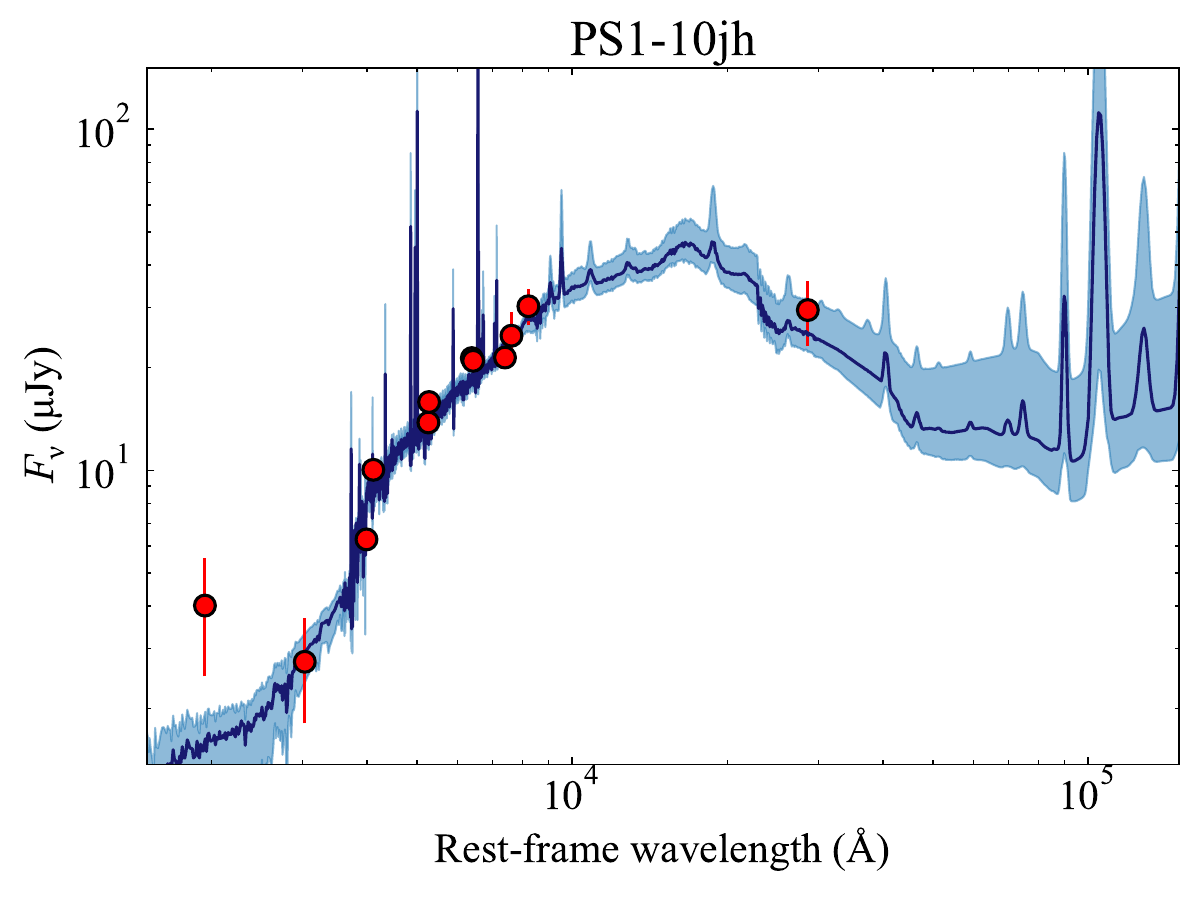}
	\includegraphics[width=0.33\textwidth]{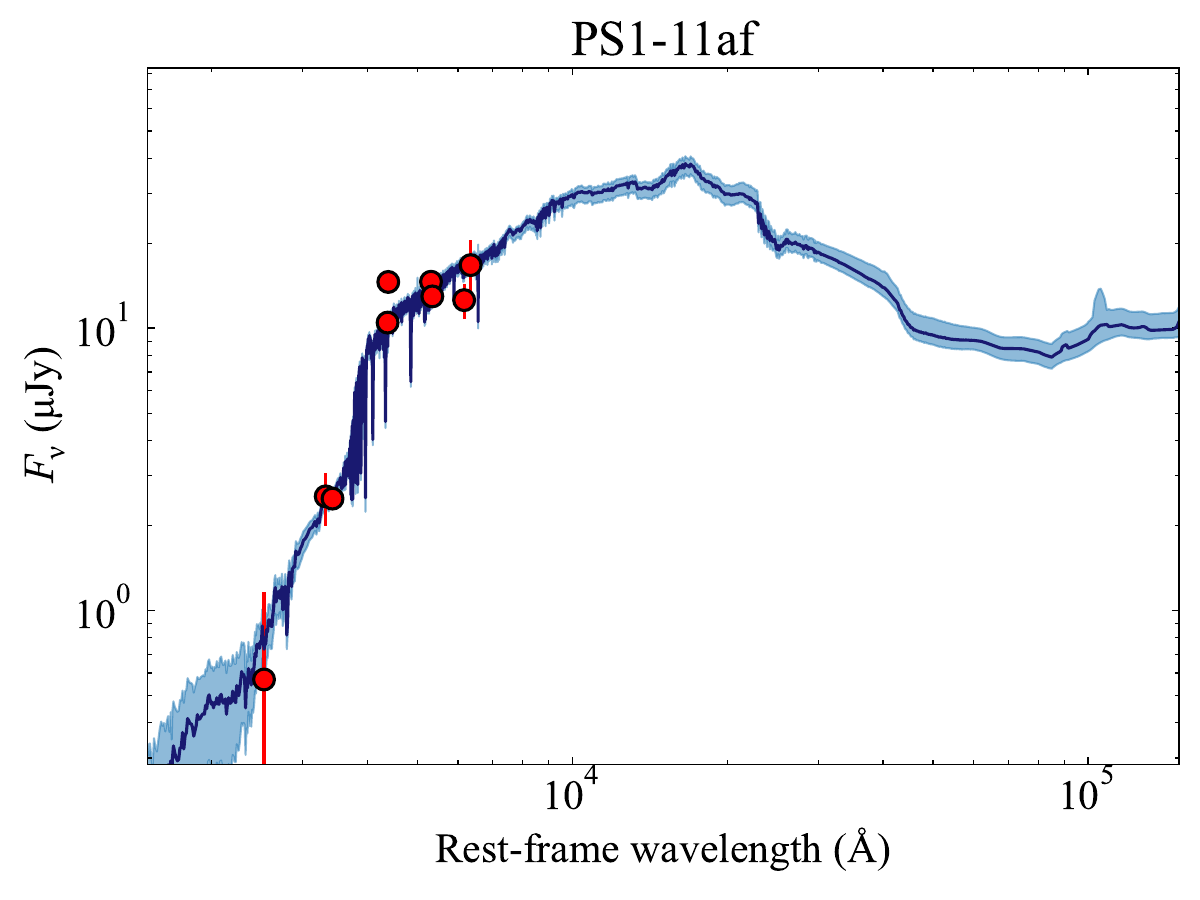}
	\includegraphics[width=0.33\textwidth]{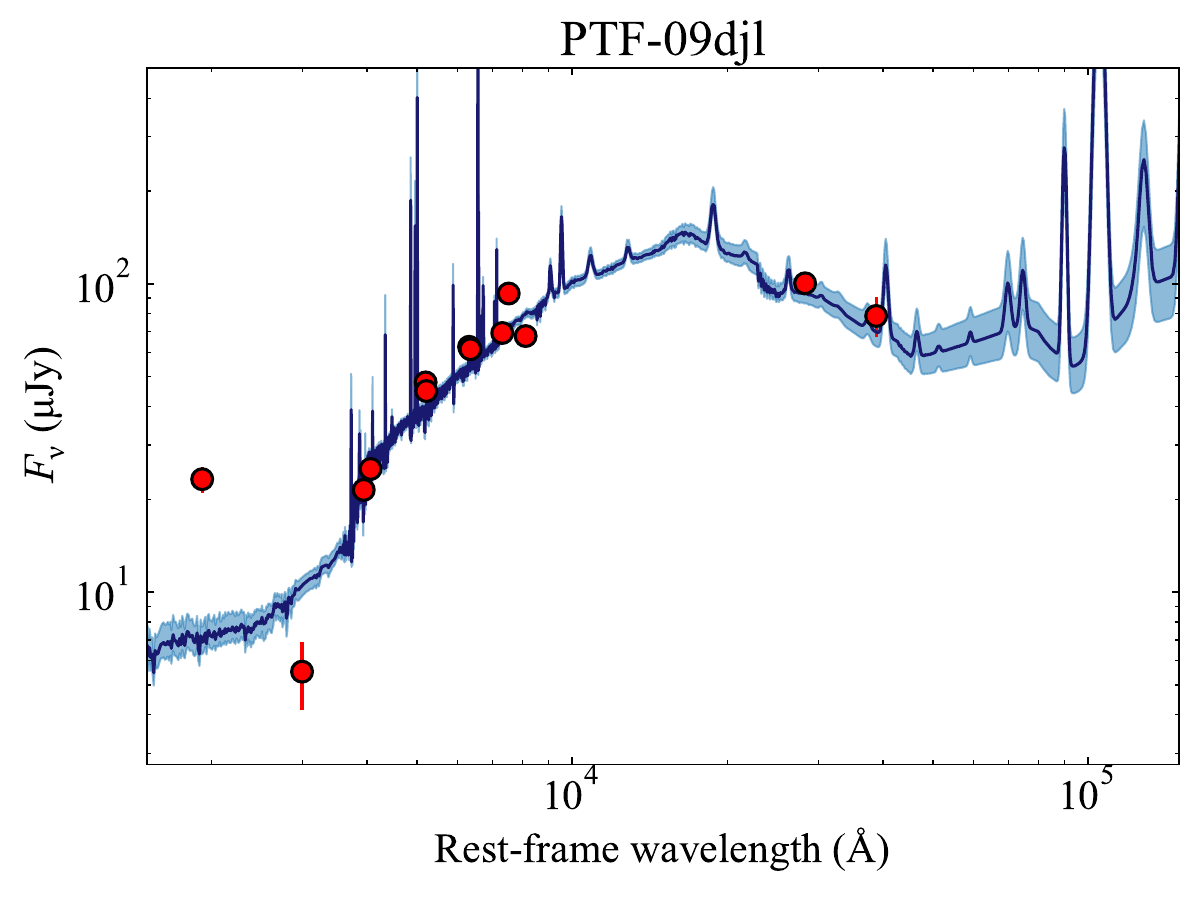}
	\includegraphics[width=0.33\textwidth]{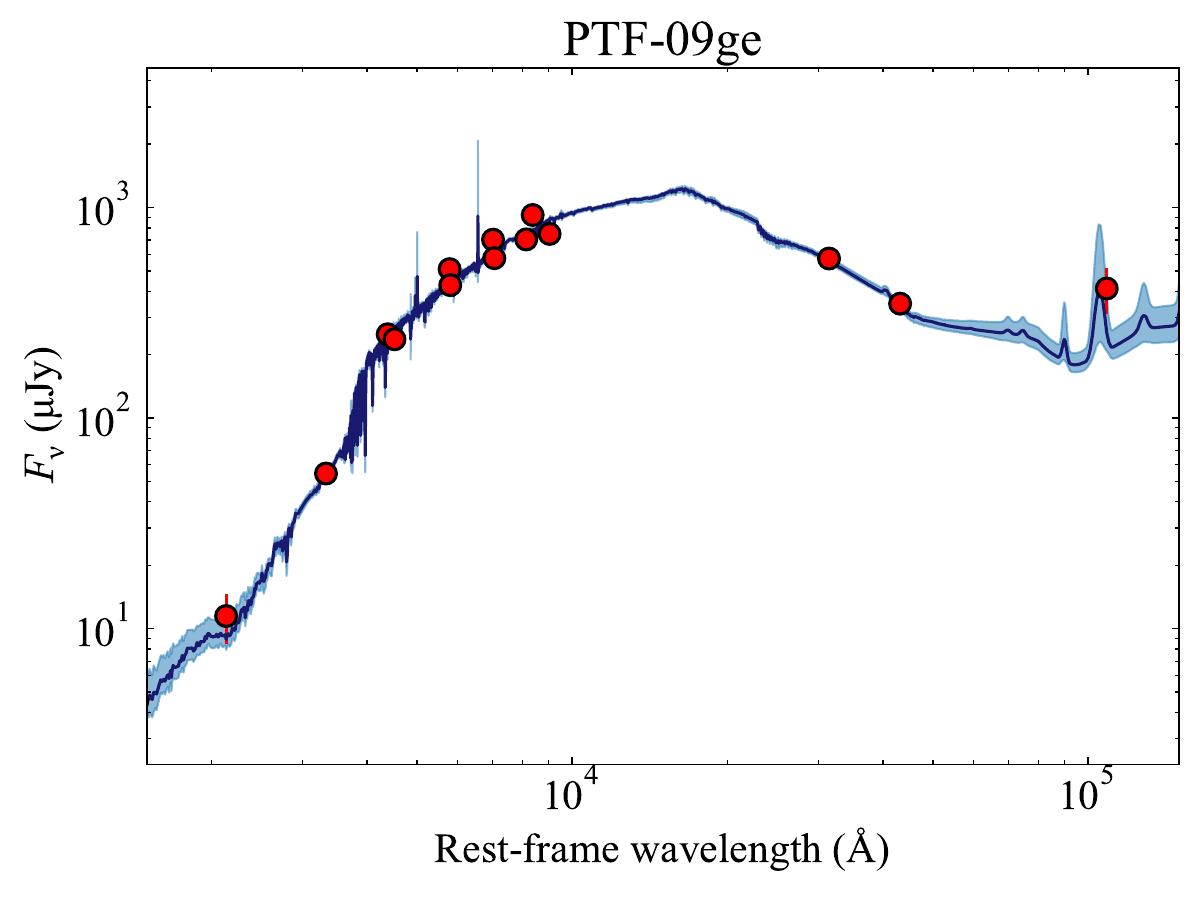}
	\includegraphics[width=0.33\textwidth]{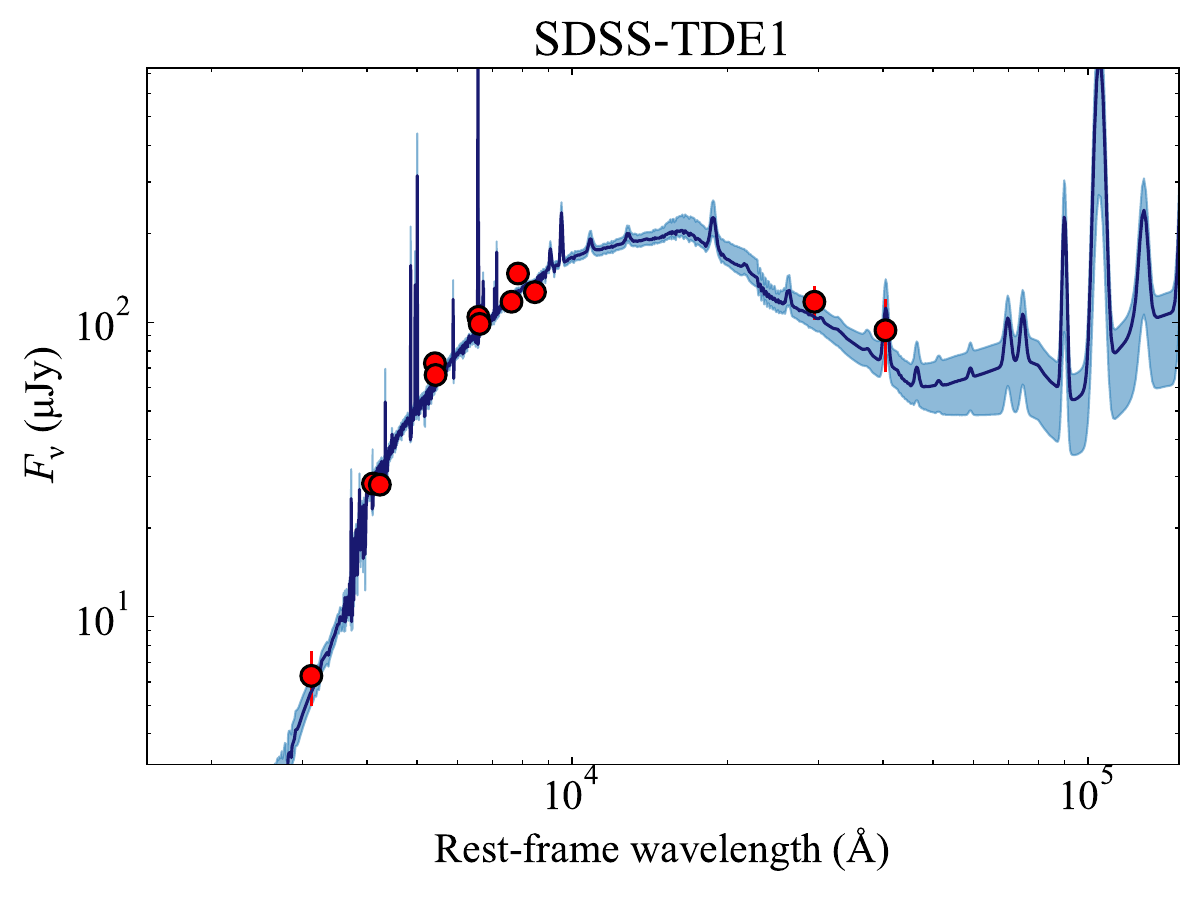}
	\includegraphics[width=0.33\textwidth]{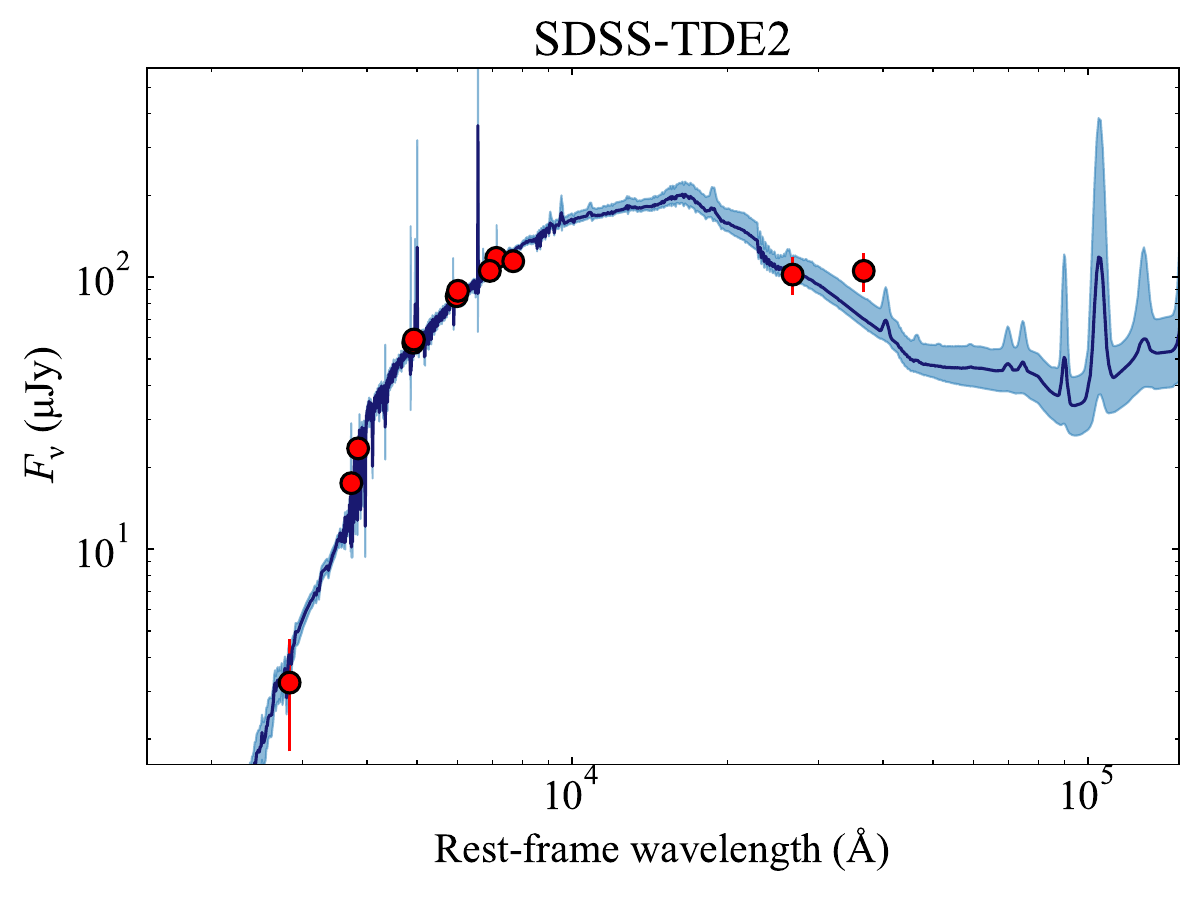}
	\includegraphics[width=0.33\textwidth]{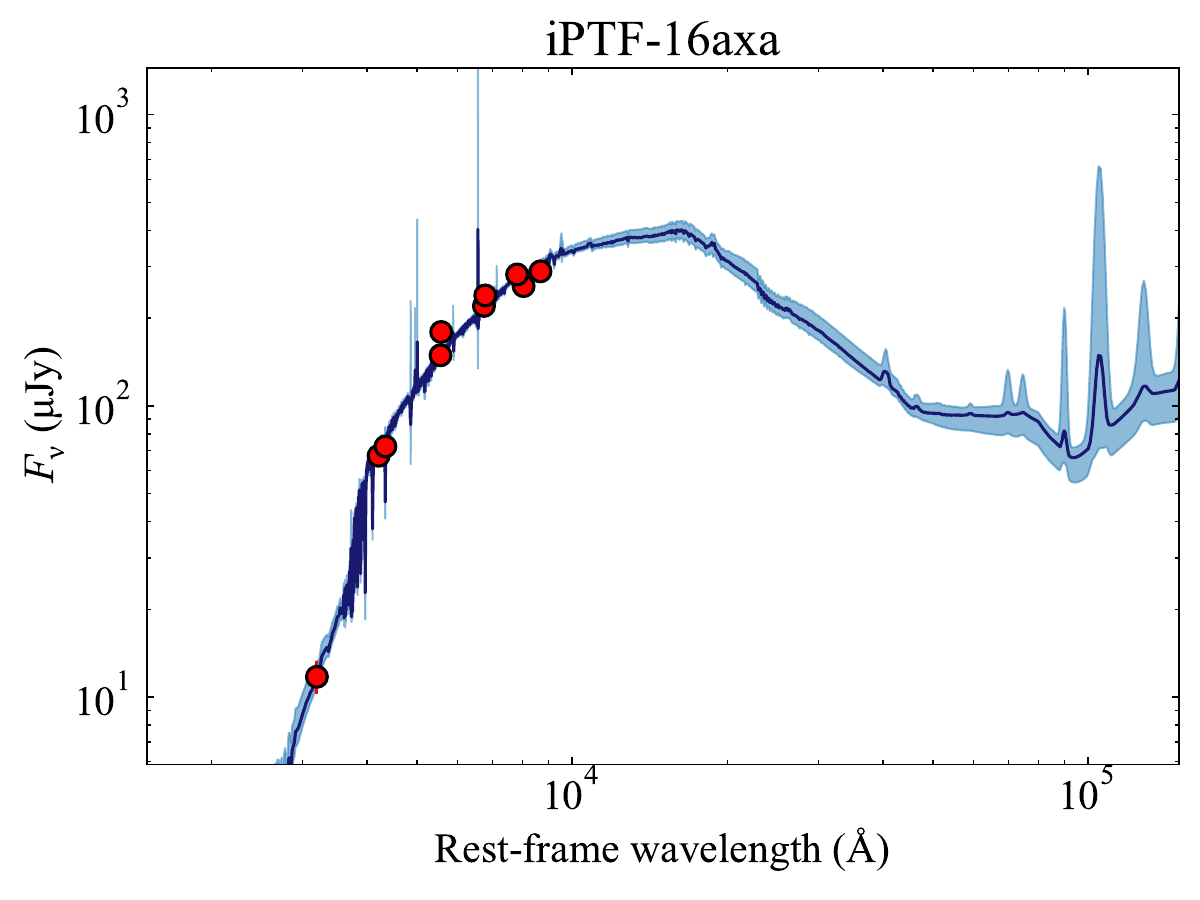}
	\includegraphics[width=0.33\textwidth]{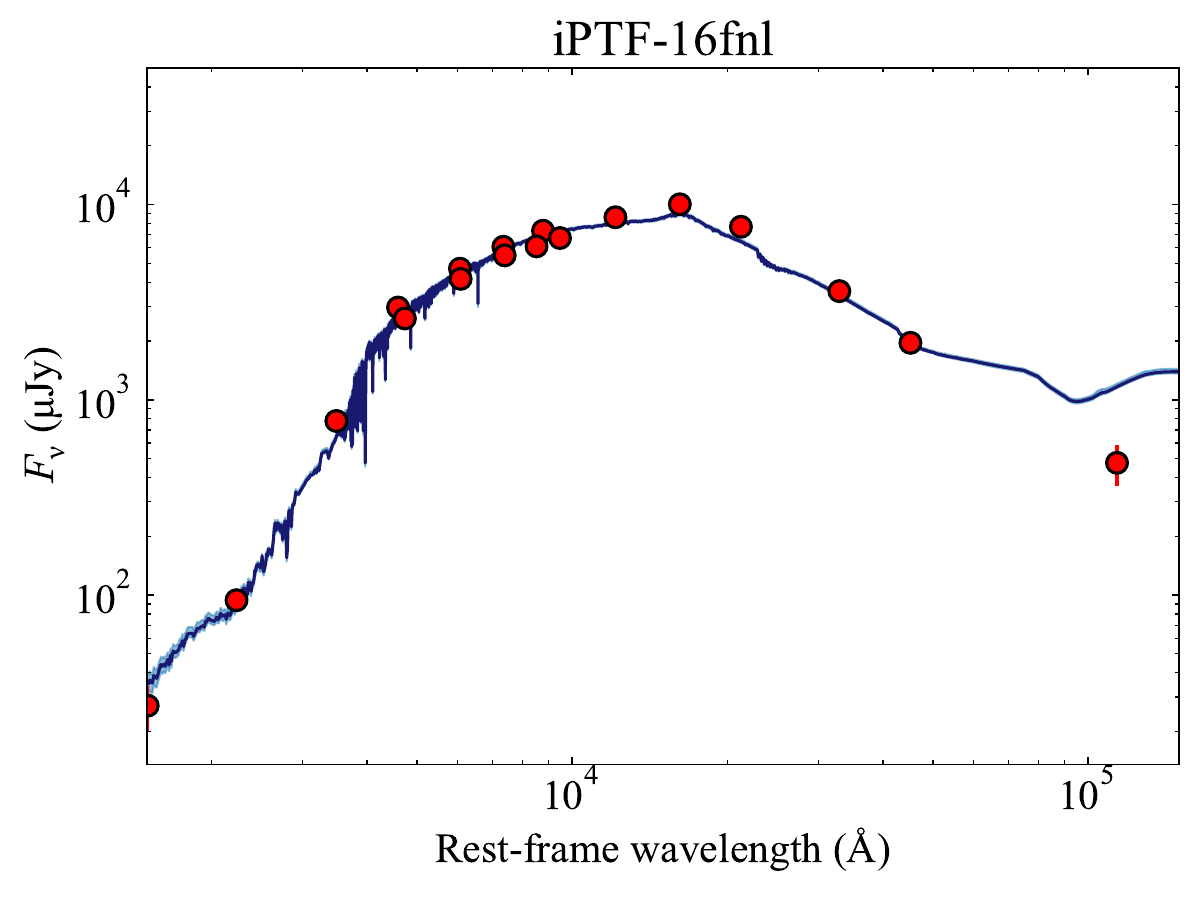}
	\caption{Archival photometry of each host galaxy (red) and the best fit model (dark blue) with the 16th to 84th percentiles of the distribution of model draws from the posterior (shaded region), derived through \textsc{Prospector}.}
    \label{fig:sed2}
\end{figure*}

Figures \ref{fig:sed1} and \ref{fig:sed2} show the \textsc{Prospector} model fits produced for each TDE host galaxy.

\section{Appendix B: Parameters}
\label{a:B}

Table \ref{t:big} lists our full sample of host galaxies, their bulge to total mass ratios and posteriors of the \textsc{Prospector} model.

\begin{landscape}

\begin{table}
\caption{Full TDE sample, with posterior medians and $1\sigma$ uncertainties from the \textsc{Prospector} model described in section \ref{s:sed} as well as B/T correction ratios discussed in section \ref{f:bulgecomp}}
  \footnotesize
  \centering
\begin{tabular}{ccccccccccccc}
\hline\hline
TDE & $\log(M_{\rm total}/{\rm M}_{\odot})$ & (B/T)g & $\log(Z/{\rm Z}_{\odot})$ & \vtop{\hbox{\strut SFH 1}\hbox{\strut 0-100 Myr}} & \vtop{\hbox{\strut SFH 2}\hbox{\strut 100-300 Myr}} & \vtop{\hbox{\strut SFH 3}\hbox{\strut 300 Myr-1 Gyr}} & \vtop{\hbox{\strut SFH 4}\hbox{\strut 1-3 Gyr}} & \vtop{\hbox{\strut SFH 5}\hbox{\strut 3-6 Gyr}} & \vtop{\hbox{\strut SFH 6}\hbox{\strut 6-13.6 Gyr}} & dust2  & dust i & dust f \\
\hline
ASASSN-14ae	& 9.94$^{+0.15}_{-0.17}$ & 0.506$\pm 0.070$ & -1.44$^{+0.33}_{-0.39}$ & -10.25$^{+0.19}_{-0.33}$ & 0.19$^{+0.31}_{-0.34}$ & 0.14$^{+0.29}_{-0.29}$ & 0.12$^{+0.32}_{-0.27}$ & 0.11$^{+0.32}_{-0.29}$ & 0.07$^{+0.31}_{-0.31}$ & 0.62$^{+0.10}_{-0.07}$ & -0.96$^{+0.05}_{-0.03}$ & 0.84$^{+0.32}_{-0.32}$	\\
ASASSN-14li & 9.78$^{+0.07}_{-0.07}$ & 0.999$\pm 0.027$ & 0.01$^{+0.09}_{-0.12}$ & -12.50$^{+0.91}_{-0.71}$ & 0.08$^{+0.22}_{-0.25}$ & 0.17$^{+0.25}_{-0.24}$ & 0.13$^{+0.29}_{-0.31}$ & 0.10$^{+0.25}_{-0.27}$ & 0.09$^{+0.32}_{-0.31}$ & 0.31$^{+0.13}_{-0.11}$ & -0.57$^{+0.15}_{-0.19}$ & 1.02$^{+0.29}_{-0.29}$	\\
ASASSN-15oi	& 9.25$^{+0.10}_{-0.08}$ & 0.454$\pm 0.073$ & 0.01$^{+0.12}_{-0.20}$ & -12.22$^{+0.99}_{-0.88}$ & 0.11$^{+0.26}_{-0.29}$ & 0.23$^{+0.25}_{-0.29}$ & 0.16$^{+0.30}_{-0.29}$ & 0.10$^{+0.27}_{-0.26}$ & 0.01$^{+0.34}_{-0.31}$ & 0.60$^{+0.16}_{-0.12}$ & -0.79$^{+0.19}_{-0.14}$ & 1.00$^{+0.30}_{-0.31}$	\\
AT2017eqx & 9.56$^{+0.11}_{-0.14}$ & 0.890$\pm 0.039$ & -1.53$^{+0.46}_{-0.33}$ & -10.57$^{+0.52}_{-1.07}$ & -0.18$^{+0.27}_{-0.24}$ & -0.04$^{+0.30}_{-0.29}$ & -0.03$^{+0.31}_{-0.29}$ & 0.02$^{+0.31}_{-0.27}$ & -0.02$^{+0.29}_{-0.31}$ & 0.48$^{+0.15}_{-0.13}$ & -0.80$^{+0.29}_{-0.14}$ & 0.99$^{+0.32}_{-0.29}$	\\
AT2018hco & 10.18$^{+0.09}_{-0.10}$ & 0.643$\pm 0.034$ & -1.30$^{+0.36}_{-0.45}$ & -11.77$^{+0.93}_{-1.07}$ & -0.28$^{+0.28}_{-0.27}$ & -0.16$^{+0.24}_{-0.24}$ & -0.10$^{+0.26}_{-0.26}$ & -0.05$^{+0.30}_{-0.30}$ & -0.05$^{+0.30}_{-0.30}$ & 0.47$^{+0.22}_{-0.19}$ & -0.25$^{+0.36}_{-0.39}$	& 0.99$^{+0.33}_{-0.31}$\\
AT2018hyz & 9.84$^{+0.09}_{-0.11}$ & 0.203$\pm 0.068$ & -1.80$^{+0.27}_{-0.13}$ & -11.32$^{+1.06}_{-1.36}$ & 0.18$^{+0.26}_{-0.26}$ & 0.22$^{+0.28}_{-0.29}$ & 0.18$^{+0.30}_{-0.32}$ & 0.12$^{+0.29}_{-0.28}$ & -0.03$^{+0.29}_{-0.27}$ & 0.61$^{+0.12}_{-0.10}$ & -0.95$^{+0.07}_{-0.04}$ & 1.02$^{+0.29}_{-0.31}$	\\
AT2018iih & 10.70$^{+0.10}_{-0.15}$ & 0.690$\pm 0.107$ & -1.47$^{+0.72}_{-0.40}$ & -9.62$^{+0.30}_{-0.28}$ & 0.06$^{+0.26}_{-0.31}$ & 0.04$^{+0.27}_{-0.26}$ & 0.04$^{+0.27}_{-0.25}$ & 0.01$^{+0.28}_{-0.29}$ & 0.02$^{+0.29}_{-0.32}$ & 0.96$^{+0.11}_{-0.09}$ & -0.94$^{+0.09}_{-0.04}$ & 0.86$^{+0.26}_{-0.28}$	\\
AT2018lna & 9.64$^{+0.13}_{-0.14}$ & 0.521$\pm 0.074$ & -1.31$^{+0.52}_{-0.46}$ & -11.26$^{+1.33}_{-1.58}$ & -0.03$^{+0.28}_{-0.29}$ & -0.00$^{+0.31}_{-0.30}$ & -0.02$^{+0.30}_{-0.32}$ & -0.04$^{+0.32}_{-0.32}$ & -0.04$^{+0.30}_{-0.31}$ & 0.64$^{+0.25}_{-0.23}$ & -0.64$^{+0.50}_{-0.27}$ & 1.05$^{+0.29}_{-0.27}$	\\
AT2018zr & 9.96$^{+0.10}_{-0.10}$ & 0.973$\pm 0.023$ & 0.04$^{+0.12}_{-0.45}$ & -11.54$^{+0.99}_{-1.06}$ & 0.08$^{+0.27}_{-0.28}$ & 0.21$^{+0.26}_{-0.29}$ & 0.15$^{+0.29}_{-0.29}$ & 0.14$^{+0.29}_{-0.30}$ & 0.05$^{+0.30}_{-0.31}$ & 0.71$^{+0.17}_{-0.16}$ & -0.78$^{+0.38}_{-0.16}$	& 1.00$^{+0.28}_{-0.32}$ \\
AT2019azh & 10.21$^{+0.08}_{-0.07}$ & 0.798$\pm 0.051$ & -0.79$^{+0.10}_{-0.12}$ & -12.64$^{+0.62}_{-0.59}$ & 0.03$^{+0.24}_{-0.21}$ & 0.07$^{+0.25}_{-0.28}$ & 0.12$^{+0.28}_{-0.31}$ & 0.07$^{+0.32}_{-0.26}$ & 0.10$^{+0.29}_{-0.29}$ & 0.34$^{+0.04}_{-0.03}$ & 0.96$^{+0.05}_{-0.03}$ & 1.00$^{+0.30}_{-0.29}$	\\
AT2019bhf & 10.56$^{+0.12}_{-0.15}$ & 0.529$\pm 0.043$ & -0.83$^{+0.36}_{-0.55}$ & -10.68$^{+0.49}_{-0.93}$ & -0.04$^{+0.27}_{-0.31}$ & -0.05$^{+0.30}_{-0.31}$ & -0.08$^{+0.28}_{-0.28}$ & -0.04$^{+0.33}_{-0.31}$ & -0.07$^{+0.31}_{-0.26}$ & 0.73$^{+0.22}_{-0.23}$ & -0.29$^{+0.20}_{-0.27}$ & 1.02$^{+0.29}_{-0.34}$	\\
AT2019cho & 10.04$^{+0.13}_{-0.20}$ & 0.676$\pm 0.090$ & -1.07$^{+0.56}_{-0.60}$ & -9.63$^{+0.40}_{-0.31}$ & -0.05$^{+0.26}_{-0.28}$ & -0.00$^{+0.32}_{-0.34}$ & 0.04$^{+0.29}_{-0.29}$ & 0.03$^{+0.30}_{-0.32}$ & 0.04$^{+0.28}_{-0.31}$ & 0.69$^{+0.12}_{-0.11}$ & -0.88$^{+0.17}_{-0.09}$ & 0.88$^{+0.30}_{-0.26}$	\\
AT2019dsg & 10.64$^{+0.10}_{-0.11}$ & 0.271$\pm 0.074$ & -1.10$^{+0.33}_{-0.45}$ & -10.47$^{+0.18}_{-0.22}$ & -0.10$^{+0.28}_{-0.28}$ & -0.11$^{+0.29}_{-0.28}$ & -0.08$^{+0.27}_{-0.28}$ & -0.04$^{+0.31}_{-0.30}$ & -0.01$^{+0.31}_{-0.28}$ & 0.77$^{+0.16}_{-0.17}$ & -0.55$^{+0.17}_{-0.22}$ & 1.00$^{+0.30}_{-0.28}$	\\
AT2019ehz & 9.90$^{+0.12}_{-0.17}$ & 0.676$\pm 0.057$ & -1.03$^{+0.42}_{-0.59}$ & -11.51$^{+1.02}_{-1.25}$ & -0.08$^{+0.27}_{-0.28}$ & -0.03$^{+0.31}_{-0.25}$ & -0.03$^{+0.28}_{-0.29}$ & -0.03$^{+0.33}_{-0.28}$ & 0.02$^{+0.30}_{-0.31}$ & 0.67$^{+0.26}_{-0.21}$ & 0.01$^{+0.22}_{-0.35}$ & 0.99$^{+0.28}_{-0.30}$	\\
AT2019eve & 9.31$^{+0.09}_{-0.10}$ & 0.864$\pm 0.053$ & -0.82$^{+0.48}_{-0.76}$ & -11.00$^{+0.63}_{-1.04}$ & -0.00$^{+0.30}_{-0.28}$ & -0.09$^{+0.32}_{-0.29}$ & -0.03$^{+0.30}_{-0.30}$ & -0.02$^{+0.28}_{-0.27}$ & -0.03$^{+0.32}_{-0.32}$ & 0.51$^{+0.34}_{-0.31}$ & -0.16$^{+0.29}_{-0.44}$ & 1.00$^{+0.31}_{-0.30}$	\\
AT2019lwu & 10.09$^{+0.09}_{-0.09}$ & 0.312$\pm 0.021$ & 0.06$^{+0.09}_{-0.17}$ & -11.84$^{+1.18}_{-1.10}$ & 0.28$^{+0.26}_{-0.27}$ & 0.26$^{+0.26}_{-0.27}$ & 0.16$^{+0.26}_{-0.31}$ & 0.19$^{+0.32}_{-0.31}$ & 0.01$^{+0.33}_{-0.28}$ & 1.46$^{+0.16}_{-0.15}$ & 0.35$^{+0.04}_{-0.06}$ & 1.02$^{+0.29}_{-0.33}$	\\
AT2019meg & 10.16$^{+0.12}_{-0.27}$ & 0.960$\pm 0.042$ & -0.23$^{+0.35}_{-0.60}$ & -9.73$^{+0.55}_{-0.79}$ & -0.01$^{+0.25}_{-0.28}$ & -0.01$^{+0.31}_{-0.30}$ & 0.02$^{+0.29}_{-0.28}$ & 0.01$^{+0.29}_{-0.26}$ & 0.04$^{+0.31}_{-0.31}$ & 1.22$^{+0.26}_{-0.22}$ & 0.25$^{+0.11}_{-0.40}$ & 0.84$^{+0.32}_{-0.32}$ \\
AT2019mha & 10.00$^{+0.08}_{-0.08}$ & 0.564$\pm 0.068$ & 0.12$^{+0.05}_{-0.08}$ & -11.79$^{+1.04}_{-1.15}$ & 0.10$^{+0.24}_{-0.24}$ & 0.30$^{+0.27}_{-0.27}$ & 0.30$^{+0.27}_{-0.29}$ & 0.17$^{+0.33}_{-0.29}$ & 0.08$^{+0.33}_{-0.30}$ & 0.48$^{+0.11}_{-0.10}$ & -0.86$^{+0.15}_{-0.10}$	& 1.01$^{+0.33}_{-0.31}$ \\
AT2019qiz & 10.27$^{+0.10}_{-0.12}$ & 0.160$\pm 0.084$ & -0.90$^{+0.26}_{-0.35}$ & -11.04$^{+0.20}_{-0.30}$ & -0.13$^{+0.27}_{-0.29}$ & -0.08$^{+0.27}_{-0.27}$ & -0.06$^{+0.29}_{-0.26}$ & -0.03$^{+0.29}_{-0.26}$ & -0.03$^{+0.28}_{-0.28}$ & 0.59$^{+0.17}_{-0.15}$ & -0.17$^{+0.21}_{-0.28}$ & 1.01$^{+0.32}_{-0.30}$	\\
D1-9 & 10.50$^{+0.11}_{-0.11}$ & 0.786$\pm 0.023$ & -0.54$^{+0.41}_{-0.68}$ & -12.15$^{+1.05}_{-0.93}$ & 0.18$^{+0.29}_{-0.26}$ & -0.12$^{+0.27}_{-0.26}$ & -0.05$^{+0.28}_{-0.33}$ & -0.07$^{+0.31}_{-0.26}$ & -0.00$^{+0.29}_{-0.29}$ & 0.54$^{+0.35}_{-0.25}$ & -0.27$^{+0.42}_{-0.44}$ & 0.97$^{+0.33}_{-0.30}$	\\
D3-13 & 11.02$^{+0.17}_{-0.16}$ & 0.941$\pm 0.018$ & -0.94$^{+0.61}_{-0.61}$ & -11.74$^{+1.05}_{-1.17}$ & -0.05$^{+0.30}_{-0.30}$ & 0.00$^{+0.29}_{-0.28}$ & -0.10$^{+0.29}_{-0.27}$ & -0.00$^{+0.28}_{-0.28}$ & -0.04$^{+0.27}_{-0.33}$ & 1.27$^{+0.44}_{-0.41}$ & 0.06$^{+0.23}_{-0.29}$ & 1.01$^{+0.28}_{-0.30}$	\\
PS1-10jh & 9.68$^{+0.12}_{-0.18}$ & 0.904$\pm 0.015$ & -0.34$^{+0.37}_{-0.69}$ & -9.95$^{+0.56}_{-0.74}$ & -0.01$^{+0.25}_{-0.27}$ & -0.02$^{+0.25}_{-0.27}$ & -0.04$^{+0.29}_{-0.32}$ & 0.02$^{+0.28}_{-0.35}$ & -0.04$^{+0.29}_{-0.31}$ & 0.61$^{+0.34}_{-0.29}$ & 0.05$^{+0.24}_{-0.38}$ & 1.00$^{+0.28}_{-0.33}$	\\
PS1-11af & 10.16$^{+0.16}_{-0.13}$ & 0.806$\pm 0.061$ & 0.07$^{+0.09}_{-0.15}$ & -12.06$^{+0.95}_{-0.91}$ & 0.08$^{+0.29}_{-0.29}$ & 0.24$^{+0.27}_{-0.35}$ & 0.22$^{+0.34}_{-0.34}$ & 0.11$^{+0.37}_{-0.33}$ & 0.08$^{+0.32}_{-0.36}$ & 0.43$^{+0.15}_{-0.13}$ & -0.80$^{+0.31}_{-0.15}$ & 0.99$^{+0.28}_{-0.27}$	\\
PTF09djl & 10.36$^{+0.13}_{-0.15}$ & 0.892$\pm 0.044$ & -0.69$^{+0.33}_{-0.81}$ & -9.62$^{+0.20}_{-0.21}$ & -0.13$^{+0.27}_{-0.26}$ & -0.11$^{+0.27}_{-0.27}$ & -0.04$^{+0.27}_{-0.30}$ & -0.09$^{+0.30}_{-0.27}$ & -0.04$^{+0.32}_{-0.30}$ & 1.03$^{+0.22}_{-0.21}$ & 0.35$^{+0.04}_{-0.06}$ & 1.05$^{+0.33}_{-0.33}$	\\
PTF09ge & 10.33$^{+0.10}_{-0.10}$ & 0.566$\pm 0.101$ & -1.06$^{+0.45}_{-0.57}$ & -11.12$^{+0.52}_{-1.06}$ & 0.00$^{+0.28}_{-0.28}$ & -0.05$^{+0.25}_{-0.28}$ & 0.02$^{+0.26}_{-0.29}$ & -0.04$^{+0.29}_{-0.26}$ & 0.01$^{+0.29}_{-0.27}$ & 0.80$^{+0.23}_{-0.25}$ & -0.29$^{+0.12}_{-0.22}$ & 0.99$^{+0.30}_{-0.28}$ \\
TDE1 & 10.21$^{+0.12}_{-0.19}$ & 0.853$\pm 0.104$ & -0.94$^{+0.42}_{-0.60}$ & -10.10$^{+0.43}_{-0.60}$ & -0.08$^{+0.27}_{-0.26}$ & -0.09$^{+0.37}_{-0.33}$ & -0.01$^{+0.31}_{-0.29}$ & -0.03$^{+0.29}_{-0.28}$ & 0.02$^{+0.28}_{-0.31}$ & 0.82$^{+0.21}_{-0.19}$ & -0.74$^{+0.20}_{-0.18}$ & 0.96$^{+0.33}_{-0.28}$	\\
TDE2 & 10.66$^{+0.12}_{-0.16}$ & 0.546$\pm 0.030$ & -0.67$^{+0.45}_{-0.49}$ & -11.34$^{+1.10}_{-1.46}$ & -0.00$^{+0.27}_{-0.29}$ & 0.01$^{+0.27}_{-0.29}$ & 0.05$^{+0.32}_{-0.28}$ & 0.00$^{+0.33}_{-0.34}$ & 0.06$^{+0.26}_{-0.30}$ & 0.61$^{+0.16}_{-0.18}$ & -0.84$^{+0.18}_{-0.11}$ & 0.98$^{+0.31}_{-0.29}$	\\
iPTF16axa & 10.31$^{+0.11}_{-0.11}$ & 0.575$\pm 0.045$ & -1.60$^{+0.46}_{-0.28}$ & -11.39$^{+0.98}_{-1.29}$ & -0.03$^{+0.30}_{-0.30}$ & 0.01$^{+0.26}_{-0.27}$ & -0.04$^{+0.32}_{-0.27}$ & 0.08$^{+0.25}_{-0.29}$ & 0.04$^{+0.28}_{-0.27}$ & 0.80$^{+0.16}_{-0.15}$ & -0.90$^{+0.16}_{-0.07}$ & 1.01$^{+0.29}_{-0.33}$	\\
iPTF16fnl & 10.12$^{+0.05}_{-0.07}$ & 0.486$\pm 0.051$ & -1.02$^{+0.12}_{-0.12}$ & -13.14$^{+0.48}_{-0.26}$ & -0.52$^{+0.22}_{-0.24}$ & -0.26$^{+0.25}_{-0.24}$ & -0.08$^{+0.31}_{-0.29}$ & -0.10$^{+0.31}_{-0.27}$ & -0.02$^{+0.27}_{-0.28}$ & 0.23$^{+0.03}_{-0.03}$ & -0.95$^{+0.08}_{-0.04}$ & 1.02$^{+0.24}_{-0.29}$	\\
\hline\hline
\end{tabular}
\\
\label{t:big}
\end{table}
\end{landscape}

% If you want to present additional material which would interrupt the flow of the main paper,
% it can be placed in an Appendix which appears after the list of references.

%%%%%%%%%%%%%%%%%%%%%%%%%%%%%%%%%%%%%%%%%%%%%%%%%%

% Don't change these lines
\bsp	% typesetting comment
\label{lastpage}
\end{document}